\documentclass{emulateapj}

\usepackage{natbib}
\usepackage{amsmath}

\newcommand{\msun}{M$_{\sun}$}
\newcommand{\rsun}{R$_{\sun}$}
\newcommand{\ldl}{$\lambda/{\Delta}{\lambda}$}
\newcommand{\teff}{T$_{\rm eff}$}
\newcommand{\logg}{$\log{g}$}

\newcommand{\rv}{$RV$}
\newcommand{\vsini}{$V_{rot}\sin{i}$}

\newcommand{\kms}{km~s$^{-1}$}

\newcommand{\name}{SDSS~J080531.84+481233.0}
\newcommand{\namesh}{SDSS~J0805+4812}

\slugcomment{Submitted 26 March 2016; Accepted for publication to ApJ 17 May 2016}

\begin{document}

\title{The Orbit of the L dwarf + T dwarf Spectral Binary SDSS~J080531.84+481233.0\footnote{Some of the data presented herein were obtained at the W.M. Keck Observatory, which is operated as a scientific partnership among the California Institute of Technology, the University of California and the National Aeronautics and Space Administration. The Observatory was made possible by the generous financial support of the W.M. Keck Foundation.}}

\author{Adam J. Burgasser\altaffilmark{1},
Cullen H. Blake\altaffilmark{2},
Christopher R.\ Gelino\altaffilmark{3},
Johannes Sahlmann\altaffilmark{4},
\&
Daniella Bardalez Gagliuffi\altaffilmark{1}
}

\altaffiltext{1}{Center for Astrophysics and Space Science, University of California San Diego, La Jolla, CA, 92093, USA; aburgasser@ucsd.edu}
\altaffiltext{2}{Department of Physics and Astronomy, University of Pennsylvania, Philadelphia, PA 19104, USA}
\altaffiltext{3}{NASA Exoplanet Science Institute, Mail Code 100-22, California Institute of Technology, 770 South Wilson Avenue, Pasadena, CA 91125, USA}
\altaffiltext{4}{European Space Agency, European Space Astronomy Centre, PO Box 78, Villanueva de la Cañada, E-28691 Madrid, Spain}

\begin{abstract}
SDSS~J080531.84+481233.0 is a closely-separated, very-low-mass binary identified through combined-light spectroscopy and confirmed as an astrometric variable. Here we report four years of radial velocity monitoring observations of the system that reveal significant and periodic variability, confirming the binary nature of the source. We infer an orbital period of {2.02$\pm$0.03~yr}, a semi-major axis of {0.76$^{+0.05}_{-0.06}$~AU}, and an eccentricity of 0.46$\pm$0.05, consistent with the amplitude of astrometric variability and prior attempts to resolve the system. Folding in constraints based on the spectral types of the components (L4$\pm$0.7 and T5.5$\pm$1.1), corresponding effective temperatures, and brown dwarf evolutionary models, we further constrain the orbital inclination of this system to be nearly edge-on (90$\degr\pm19\degr$), and deduce a large system mass ratio {(M$_2$/M$_1$ = 0.86$^{+0.10}_{-0.12}$)}, substellar components {(M$_1$ = 0.057$^{+0.016}_{-0.014}$~{\msun}, M$_2$ = 0.048$^{+0.008}_{-0.010}$~{\msun})}, and a relatively old system age (minimum {age = 4.0$^{+1.9}_{-1.2}$~Gyr}).  The {measured} projected rotational velocity of the primary ({\vsini} = {34.1$\pm$0.7~{\kms}}) implies that this inactive source is a rapid rotator (period $\lesssim$ 3~hr) and a viable system for testing spin-orbit alignment in very-low-mass multiples. Robust model-independent constraints on the component masses may be possible through measurement of the reflex motion of the secondary at wavelengths in which it contributes a greater proportion of 
the combined luminence, while the system may also be resolvable through sparse-aperature mask interferometry with adaptive optics. The combination of well-determined component atmospheric properties and masses near and/or below the hydrogen minimum mass make SDSS~J0805+4812AB an important system for future tests of brown dwarf evolutionary models.
\end{abstract}


\section{Introduction}

Multiple systems, particularly short-period binaries, are key targets for fundamental measurements of individual stars.
While the orbital periods of these systems allow us to infer {their} total system mass, the gold standard is the determination of individual component masses through absolute astrometry or reflex motion from both components, and radii through transits or modeling of the spectral energy distribution. These quantities can be used to directly test stellar structure models {\citep{2006Natur.440..311S,2010A&ARv..18...67T,2011ApJ...730...79J,2013MNRAS.428.1656B}}. For brown dwarfs, objects with insufficient mass to sustain core hydrogen fusion \citep{1962AJ.....67S.579K,1963ApJ...137.1121K,1963PThPh..30..460H}, such systems provide empirical tests of evolutionary cooling models, where the combination of mass and atmospheric properties, in some cases coupled with external information on system age or composition, can be directly compared to model predictions \citep{2009ApJ...692..729D,2014ApJ...790..133D,2009ApJ...695..788K,2010ApJ...711.1087K,2010ApJ...722..311L,2012ApJ...757..110B}. In addition, given that a significant fraction (15--30\%) of very low mass (VLM; M $\leq$ 0.1~{\msun}) stars and brown dwarfs are found in $a$ $\lesssim$ 20~AU binary systems (e.g., \citealt{2007ApJ...668..492A,2007prpl.conf..427B,2013AN....334...36D,2013ARA&A..51..269D} and references therein), the orbital properties of these systems and degree of spin-orbit alignment provide necessary constraints on brown dwarf formation mechanisms, which are still under investigation \citep{2009MNRAS.392..590B,2012MNRAS.419.3115B,2010ApJ...725.1485O,2011ASPC..447...47K,2014MNRAS.442.3722P}.

Detecting resolvable VLM binaries with short enough orbital periods for mass measurement can be challenging, and just over a dozen such systems are currently known (e.g., \citealt{2011ApJ...733..122D}). Even fewer radial velocity  (e.g., \citealt{1999AJ....118.2460B,2002AJ....124..519R,2008ApJ...678L.125B,2010A&A...521A..24J,2012ApJ...757..110B}) and astrometric variables (e.g., \citealt{2008ApJ...686..548D,2012ApJS..201...19D,2013A&A...556A.133S}) are known, and in many cases the component properties of these systems cannot be resolved. {Only a single eclipsing brown dwarf-brown dwarf system has been found, a $\sim$1~Myr-old system in Orion \citep{2006Natur.440..311S}; although several brown-dwarf-mass objects have been found to transit more massive stars (e.g., \citealt{2008A&A...491..889D,2011ApJ...730...79J,2014A&A...564A..98S}), enabling radii measurements that are in many cases inconsistent with models (e.g., \citealt{2011ApJ...736...47B}).} Fortunately, the spectra of M-, L- and T-type brown dwarfs are sufficiently distinct that binaries composed of these sources can often be discerned and characterized through unresolved spectroscopy; these are the VLM spectral binaries \citep{2004ApJ...604L..61C,2010ApJ...710.1142B,2014ApJ...794..143B}. A dozen of these systems have been identified and confirmed over the past decade, over half of which have compact orbits ($\lesssim$2~AU) enabling simultaneous orbital mass measurements and component atmospheric characterization \citep{2015AJ....150..163B}. 

One of these systems is {\name} (hereafter {\namesh}), a peculiar L dwarf identified in the Sloan Digitial Sky Survey (SDSS \citealt{2000AJ....120.1579Y}) {that exhibits} highly divergent optical (L4; \citealt{2002AJ....123.3409H}) and near-infrared (L9.5, \citealt{2004AJ....127.3553K}) spectral classifications. It was identified as a potential L dwarf plus T dwarf binary on the basis of its spectral peculiarities \citep{2007AJ....134.1330B}, and found to be an astrometric variable by
\citet{2012ApJS..201...19D} with an amplitude of $\approx$15~mas.
While unable to constrain the orbit of the system, \citet{2012ApJS..201...19D} estimated a semi-major axis of
0.9--2.3~AU and orbital period of 2.7--9.1~yr, but found no evidence of a resolved companion
in unpublished observations with Keck Laser Guide Star Adaptive Optics (LGSAO; \citealt{2006PASP..118..310V,2006PASP..118..297W}) observations.
\citet{2015AJ....150..163B} have also reported this source as unresolved in two epochs of Keck LGSAO imaging with an angular separation upper limit of 170~mas, taking into account the expected flux ratio of the system.  

In this article, we report the detection of significant and periodic radial velocity variations in high-resolution spectroscopic monitoring of {\namesh} that allow us to make the first robust constraints on the orbital and physical properties of the system components. In Section~\ref{sec:observations} we describe the observations and data analysis methodology that yield both radial and rotational velocities for the source. 
In Section~\ref{sec:spex} we update the spectral characterization of the {\namesh}AB components through a revised
analysis of its combined-light near-infrared spectrum.
In Section~\ref{sec:orbit} we briefly describe our orbital analysis and determination of the system parameters, including constraints based on the component spectral types and evolutionary models. We discuss our results in Section~\ref{sec:discuss}. {A detailed} description of our spectral analysis and orbital modeling are provided in the Appendices.

\section{Observations and Data Analysis\label{sec:observations}}

High resolution near-infrared spectra of {\namesh} were obtained with 
the Near InfraRed SPECtrometer (NIRSPEC; \citealt{2000SPIE.4008.1048M}) on the Keck II telescope over {14 epochs between 2012 April 02 and 2016 April 22} (Table~\ref{tab:nirspec}). In all cases, data were acquired using the N7 order-sorting filter and 0$\farcs$432-wide slit to obtain 2.00--2.39~$\micron$ spectra over orders 32--38 with {\ldl} $\approx$ 20,000 ($\Delta{v}$ $\approx$ 15~{\kms}) and dispersion of 0.315~{\AA}~pixel$^{-1}$. For each observation, two exposures of 1200--1500~s each were obtained, nodding 7$\arcsec$ along the slit, followed by observations of the nearby A0~V star HD~71906 ($V$ = 6.18) at a similar airmass.  Flat field and dark frames were obtained at the start of each night for detector calibration.  

We improved upon the forward-modeling process described in \citet{2015AJ....149..104B} by incorporating an iterative, multi-threaded Markov Chain Monte Carlo (MCMC) algorithm to achieve more consistent results across observations; our methodology is detailed in Appendix~A. 
Figure~\ref{fig:nirspec} illustrates a sample extraction and fit from our 2016 February 16 (UT) observations, and the resulting radial and rotational velocities for all epochs analyzed are listed in Table~\ref{tab:nirspec}. 
The median signal-to-noise (S/N) of the extracted data in order 33 (2.29--2.33~$\micron$) ranged from 6--23. Measured
uncertainties spanned 0.3--1.8~{\kms} for radial velocities and 0.4--1.8~{\kms} for rotational velocities, in line with spectral S/N.
We infer an additional systematic uncertainty for the rotational velocities of {2.5~{\kms}} by enforcing consistency in the measurements ($P(\chi^2,DOF) < 90$\%)\footnote{$P(\chi^2,DOF)$ is the probability distribution function of the $\chi^2$ distribution for degrees of freedom (DOF) equal to the number of measurements $N$ minus one. We use the 
standard defintion of $\chi^2 = \sum_i^N\frac{(m_i-\bar{m})^2}{\sigma_i^2}$, where $m_i$ are the measured values, $\bar{m}$ the uncertainty-weighted mean, and $\sigma_i$ the measurement uncertainties.} over eleven epochs, excluding the S/N $<$ 10 data from 2015 December 8 and 2016 January 1 (UT).  The uncertainty-weighted mean rotational velocity is 34.1$\pm$0.7~{\kms}, with no significant correlation between radial and rotational velocities.
Assuming a radius of 0.084~{\rsun}, based on the evolutionary models of \citet{2003A&A...402..701B} for an effective temperature {\teff} = 1700~K and age of 4~Gyr (see below), this projected velocity translates into a maximum rotational period of {3.0}~hr. Like many mid-type L dwarfs, {\namesh}A is a rapidly rotating dwarf which nevertheless lacks magnetically-driven nonthermal H$\alpha$ emission \citep{2002AJ....123.3409H}.

\begin{figure*}
\epsscale{1.0}
\plotone{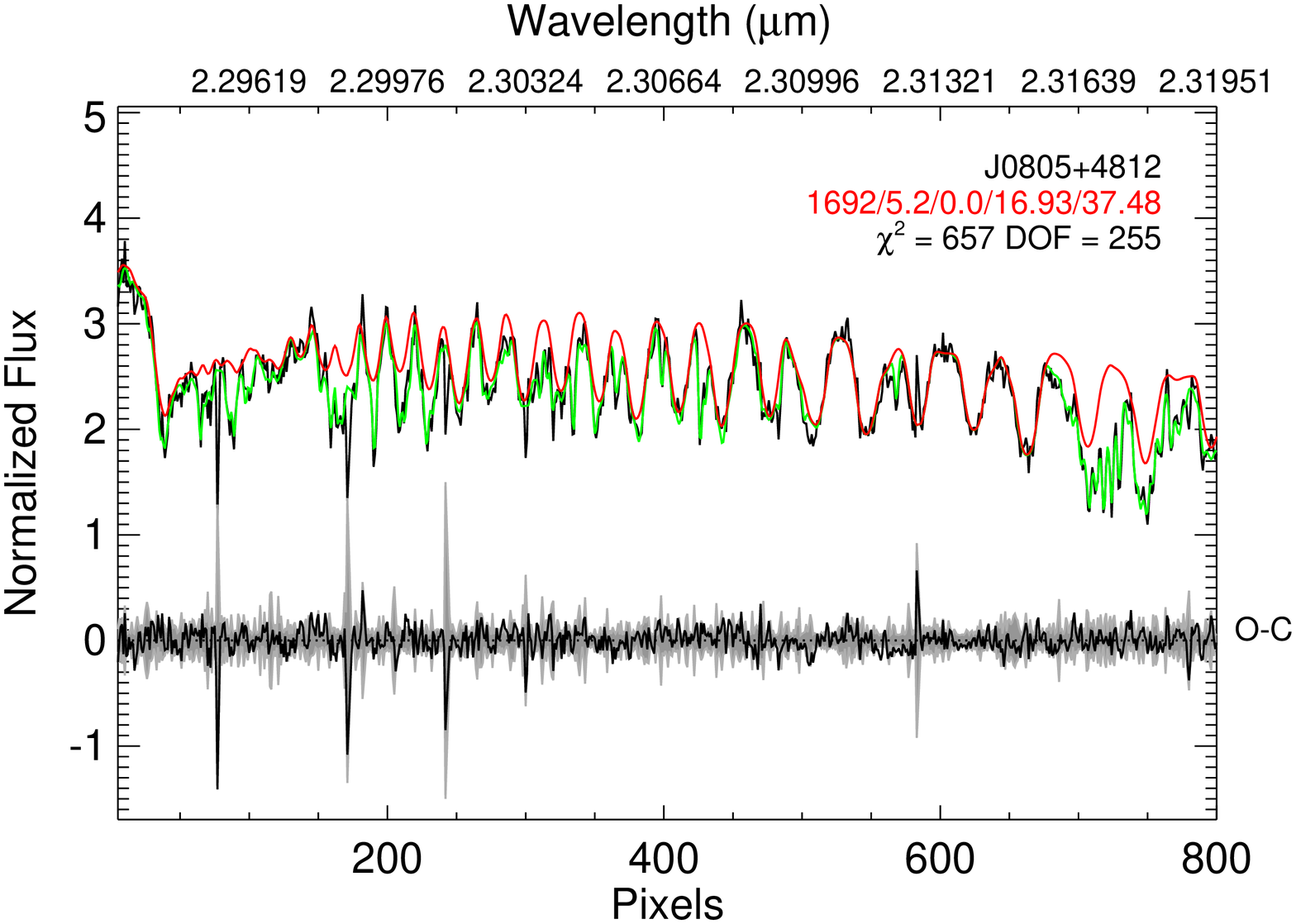}
\caption{NIRSPEC order 33 spectrum of {\namesh} obtained on UT 2016 February 16 (black line), compared to the best-fit interpolated atmosphere model from \citet[red line]{2012RSPTA.370.2765A}, parameterized as {\teff} = 1692~K, {\logg} = 5.2, [M/H] $\equiv$ 0, RV = 16.93~{\kms} and {\vsini} = 37.48~{\kms}. The best-fit model times scaled telluric absorption is shown as the green line.  Pixel scale is listed along the bottom while wavelength scale is listed along the top.
The difference between data and model (O-C)  is shown in black at the bottom of the plot; the $\pm$1$\sigma$ uncertainty spectrum is indicated in grey.  The $\chi^2$ = 657 and 255 degrees of freedom (DOF) indicate a reasonable fit.
\label{fig:nirspec}}
\end{figure*}

The radial velocities are inconsistent with a constant value {($\chi^2$ = 368, DOF = 13) and display periodic} variation. We interpret this behavior as the reflex motion of the primary under the gravitational influence of a brown dwarf secondary, and use these data to infer the orbit of the system as described below.

\section{Re-examination of the Spectral Composition of {\namesh}\label{sec:spex}}

\begin{figure}
\epsscale{1.0}
\plotone{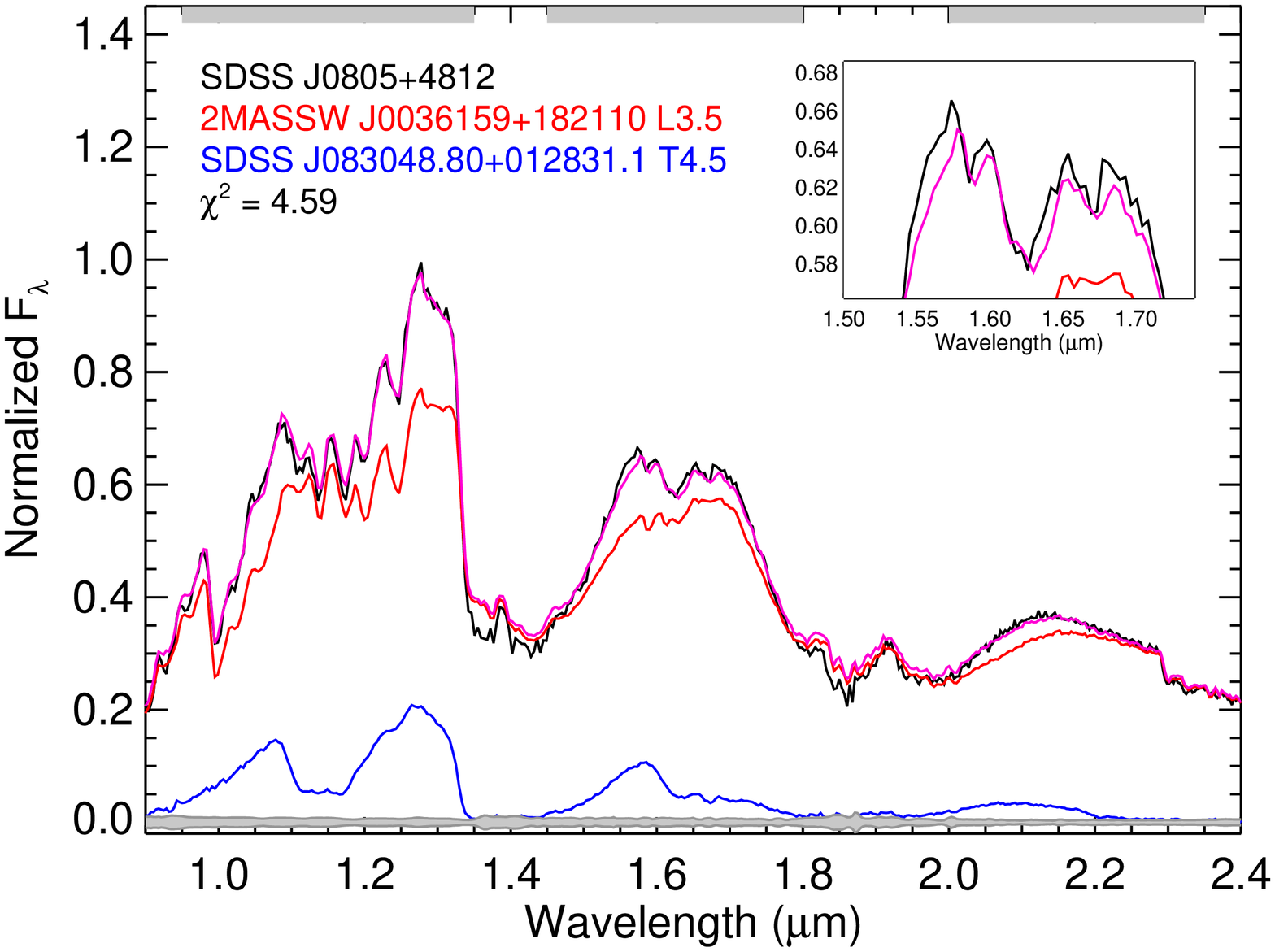}
\caption{Best-fit spectral binary template (purple line) compared to the combined-light SpeX spectrum of {\namesh} (black line). The template is composed of the L3.5 2MASS~J0036159+182110 (red linel data from \citealt{2008ApJ...681..579B}) and the T4.5 SDSS~J083048.80+012831.1 (blue line; data from \citealt{2010ApJ...710.1142B}), both shown at their relative scaling.  The gray bars at top indicate the regions over which the fitting was done.
The $\pm$1$\sigma$ uncertainty spectrum of {\namesh} is shown in gray along the bottom. The inset box highlights the 1.50--1.75~$\micron$ range where the ``dip'' feature is seen, arising from overlapping FeH and CH$_4$ absorption from pimary and secondary, respectively.}
\label{fig:binaryfit}
\end{figure}

The initial identification of {\namesh}AB as a spectral binary candidate was based on comparison of its
blended-light spectrum with 50 L and T dwarf templates. {That analysis inferred component types of L4.5 and T5.} 
\citet{2012ApJS..201...19D}, using a similar technique, inferred equivalent classifications of L5: and T5. We revisited these analyses
following the procedures described in \citet{2010ApJ...710.1142B}, comparing the SpeX spectum of {\namesh} to 534 L2--T8 spectral templates from an updated
SpeX Prism Library (SPL; \citealt{2014ASInC..11....7B}), and 76,873 binary templates constructed from these templates after scaling them to absolute magnitudes using the \citet{2012ApJS..201...19D} $M_J$/spectral type relation.  The best-fit spectral binary template, composed of  
the L3.5 2MASS~J0036159+182110 \citep{2000AJ....119..369R} and the T4.5 SDSS~J083048.80+012831.1 \citep{2004AJ....127.3553K} is shown in Figure~\ref{fig:binaryfit}. 
F-test statistic-weighted means from the best 100 fits (lowest $\chi^2$) yield decimal component types of L4.2$\pm$0.7 and T5.5$\pm$1.1, consistent with prior determinations.  We use these component types and their uncertainties in our analysis of the orbital properties of the system below.

\begin{deluxetable*}{lccclccc}
\tablecaption{NIRSPEC Observations and Measurements\label{tab:nirspec}}
\tabletypesize{\small}
\tablewidth{0pt}
\tablehead{
\colhead{UT Date} &
\colhead{MJD} &
\colhead{t$_{int}$} &
\colhead{Airmass} &
\colhead{Conditions} &
\colhead{S/N} &
\colhead{{\rv}\tablenotemark{a}} &
\colhead{\vsini\tablenotemark{a}} \\
& &
\colhead{(s)} & & & &
\colhead{({\kms})} &
\colhead{({\kms})} \\
}
\startdata
2012 Apr 02 &	56019.28665 & 2400 & 1.18 &  clear, 0$\farcs$7 & 23 & 14.7$\pm$0.3 & 38.0$\pm$0.4 \\
2012 Nov 27 & 	56258.47308 &	2400 & 1.30 & p.\ cloudy, 0$\farcs$5   & 16 & 7.6$\pm$0.5 & 37.2$\pm$0.7 \\
2013 Jan 20 & 56312.46857 &	3000 & 1.15 & clear, 1$\arcsec$  & 10 & 9.3$\pm$0.5 & 36.1$\pm$0.8 \\
2013 Feb 05 &	56328.48671 &	3000 & 1.33 & clear, 1$\arcsec$  & 12 & 7.7$\pm$0.5 & 37.0$\pm$1.2 \\
2013 Sep 17 & 	56552.62423&	2400 & 1.78  &  clear, 1--2$\arcsec$ & 12 & 7.6$\pm$0.8 &  37.3$\pm$1.8 \\
2013 Oct 16 &	56581.62182&	2500 & 1.28  & p.\ cloudy, 0$\farcs$8  & 19 & 10.8$\pm$0.3 &  37.2$\pm$0.5 \\
2014 Apr 13 &	56760.26506&	3000 & 1.15 & clear, 0$\farcs$5  & 21 & 14.4$\pm$0.4 & 35.9$\pm$0.9 \\
2014 Dec 08 &	56999.59552&	3000 & 1.14 & clear, 0$\farcs$8  & 7 & 6.9$\pm$1.5 &  26.7$\pm$0.9 \\
2015 Jan 01 &	57023.52876&	3000 & 1.16 & cloudy, 1$\arcsec$  & 6 & 9.7$\pm$1.8 & 28.4$\pm$1.7 \\
2015 Dec 29 &	57385.54397&	3000 & 1.16  & clear, 0$\farcs$5  & 22 & 17.5$\pm$0.4 & 39.0$\pm$0.7 \\
2016 Jan 18 &	57405.48029&	3000 & 1.18  & clear, 0$\farcs$5  & 16 & 17.5$\pm$0.5 &  36.1$\pm$0.9 \\
2016 Feb 03 &	57421.40203&	2800  & 1.13 &  clear, 1--2$\arcsec$ & 15 & 17.8$\pm$0.6 &  35.4$\pm$1.0 \\
2016 Feb 16 &	57434.32974&	2400  & 1.15 & clear, 0$\farcs$6  & 22 & 16.3$\pm$0.7 &  37.2$\pm$0.5 \\
2016 Apr 22 &	57500.25885&	2400  & 1.20 & clear, 0$\farcs$5  & 15 & 14.7$\pm$0.5 &  39.0$\pm$0.8 \\
\enddata
\tablenotetext{a}{Additional systematic errors of 2.5~{\kms} for these parameters are not included in the values listed here.}
\tablenotetext{b}{These values were not included in the computation of the mean {\vsini} = 34.1$\pm$0.7~{\kms}; see Section~\ref{sec:orbit}.}
\end{deluxetable*}

\section{Determining the Orbit of {\namesh}AB\label{sec:orbit}}

\begin{figure}
\epsscale{1.0}
\plotone{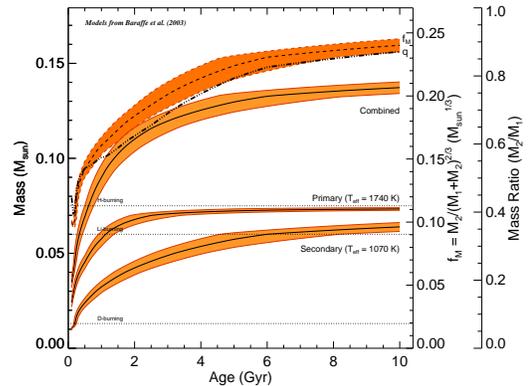}
\caption{Estimated component and combined masses of {\namesh}A and B as a function of system age
(solid lines) based on their estimated L4$\pm$0.7 and T5.5$\pm$1.1 classifications, corresponding {\teff} estimates of 1740$\pm$100~K and 1070$\pm$80~K, respectively; and the
evolutionary models of \citet{2003A&A...402..701B}. The dashed line shows the corresponding mass function,
$f_M$ = M$_2$/(M$_1$+M$_2$)$^{2/3}$ (near right axis), while the triple-dot dash line shows the mass ratio
$q$ = M$_2$/M$_1$ (far right axis). The deuterium, lithium and hydrogen burning minimum-mass limits
are labeled as dotted lines.}
\label{fig:model_masses}
\end{figure}

We analyzed the radial velocity curve of {\namesh} using an improved MCMC orbit-fitting code based on  \citet{2015AJ....149..104B} and described in detail in Appendix~B.  Two separate fits were made to the data: an ``unconstrained'' fit with a weak limit on the total mass of the system (M$_{tot}$ $\leq$ 0.3~{\msun}) and a ``constrained'' fit in
 which the orbit-deduced component and total system masses were compared to predictions from evolutionary models and component effective temperatures, following \citet{2009AJ....137.4621B}.  The temperatures were estimated from several {\teff}/spectral type relations \citep{2004AJ....127.3516G,2008ApJ...685.1183L,2009ApJ...702..154S,2013AJ....146..161M,2015ApJ...810..158F}, yielding 1650--1825~K for {\namesh}A and 990--1140~K for {\namesh}B, with uncertainties\footnote{Uncertainties include the spectral type uncertainties and systematic uncertainties in the relations.} of 100--150~K. We converged on average values of 1740$\pm$100~K and 1070$\pm$80~K.  These temperatures were converted to age-dependent masses using the solar-metallicity models of \citet{2003A&A...402..701B}. Figure~\ref{fig:model_masses} displays the component and total system masses as a function of age, as well as the mass function
\begin{equation}
f_M^{(evol)}  = \frac{M_2}{\left(M_1+M_2\right)^{2/3}}. \label{eqn:fmevol}
\end{equation}
These parameters were used to restrict solutions in the constrained orbit fit.
We determined that an additional 0.5~{\kms} systematic uncertainty in the radial velocity measurements
is required based on the $\chi^2$ of the best-fit orbit models.

\begin{figure*}
\epsscale{1.1}
\centering
\includegraphics[height=7cm]{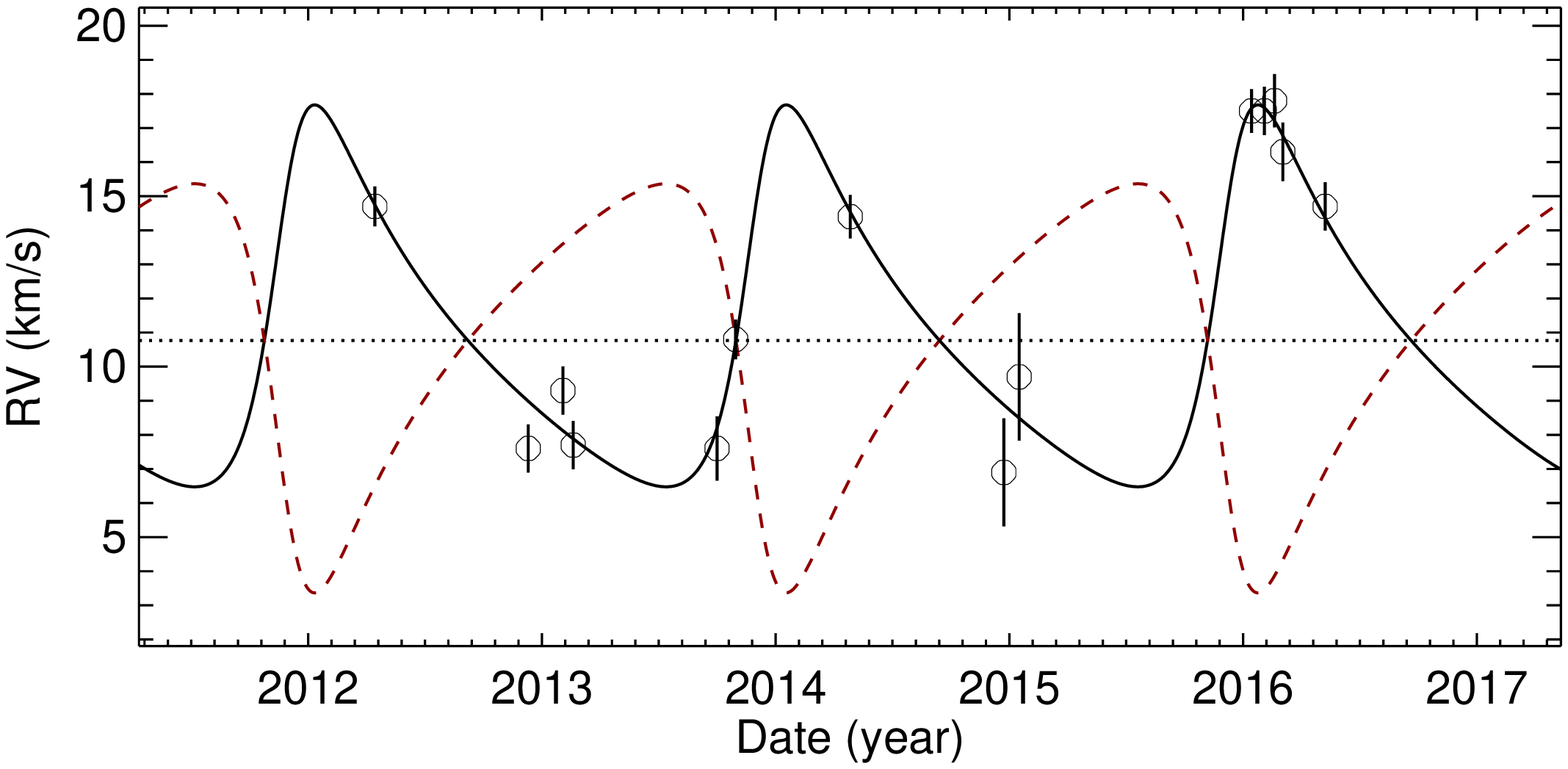} \\
\includegraphics[height=7cm]{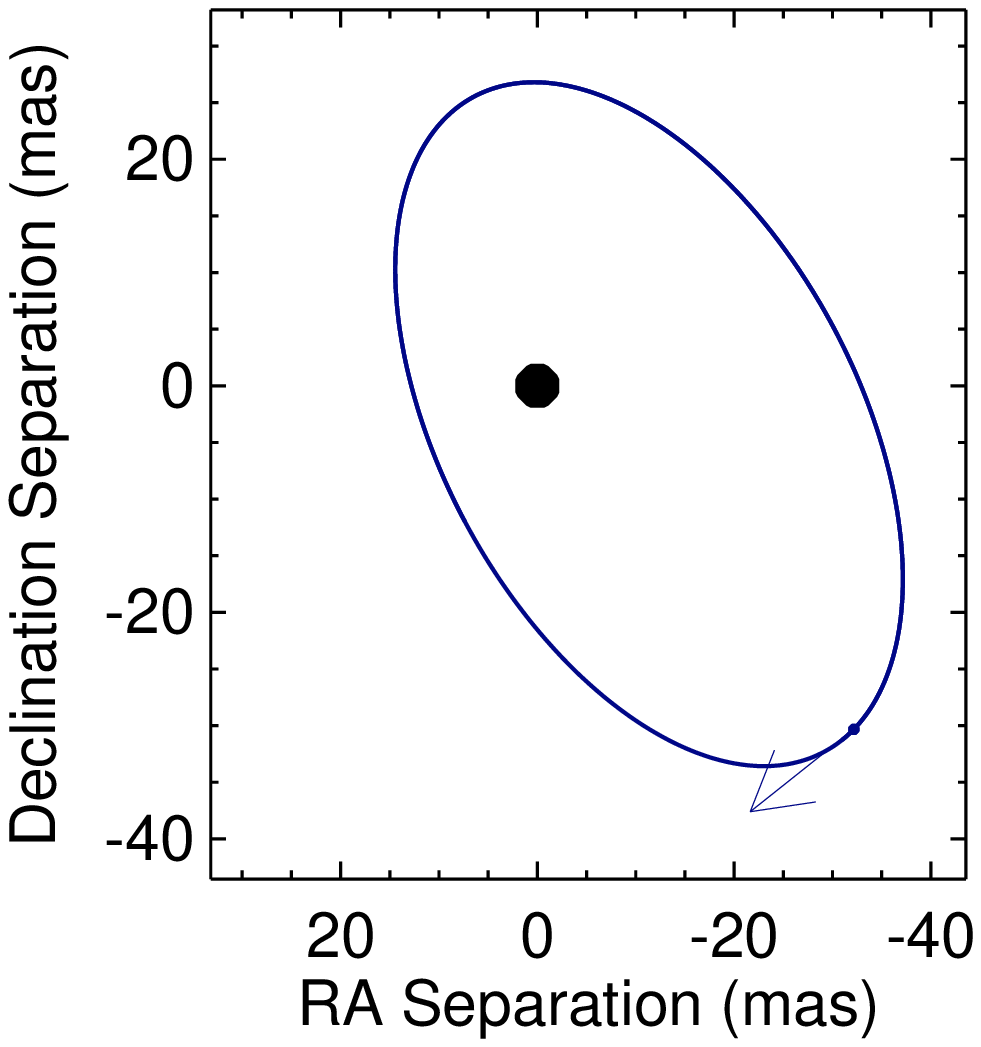} 
\includegraphics[height=7cm]{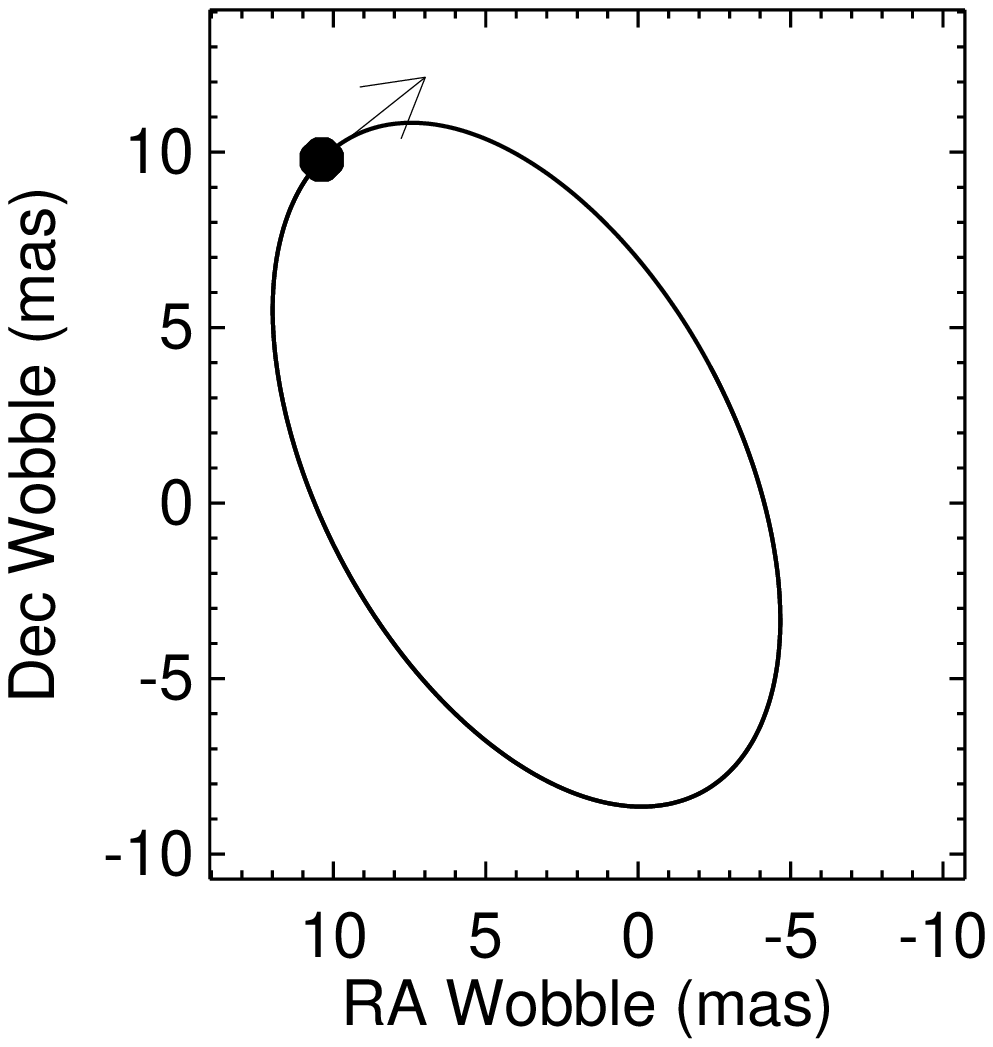} \\
\caption{Best-fit (minimum $\chi^2$) orbital solution from MCMC orbital analysis of NIRSPEC data, 
unconstrained by evolutionary models.
The top panel shows the predicted radial motion of both the primary (solid black line) and secondary (red dashed line) as compared to primary radial velocity measurements (open circles with error bars). 
The bottom left panel shows the predicted orbital motion of the secondary (blue line) relative to the primary (black dot at the origin) projected on the sky.
The bottom right panel shows the predicted astrometric orbital motion of the primary (distinguished from parallactic motion) projected on the sky. 
In the bottom panels, the arrow indicates the direction of orbital motion (secondary or primary) at apoapsis, and the orbits are shown at an arbitrary longitude of ascending node, which is unconstrained in these observations.
Parameters for these fits are listed in Table~\ref{tab:orbit_parameters}.
\label{fig:orbit}}
\end{figure*}

\begin{figure*}
\epsscale{1.1}
\centering
\includegraphics[height=7cm]{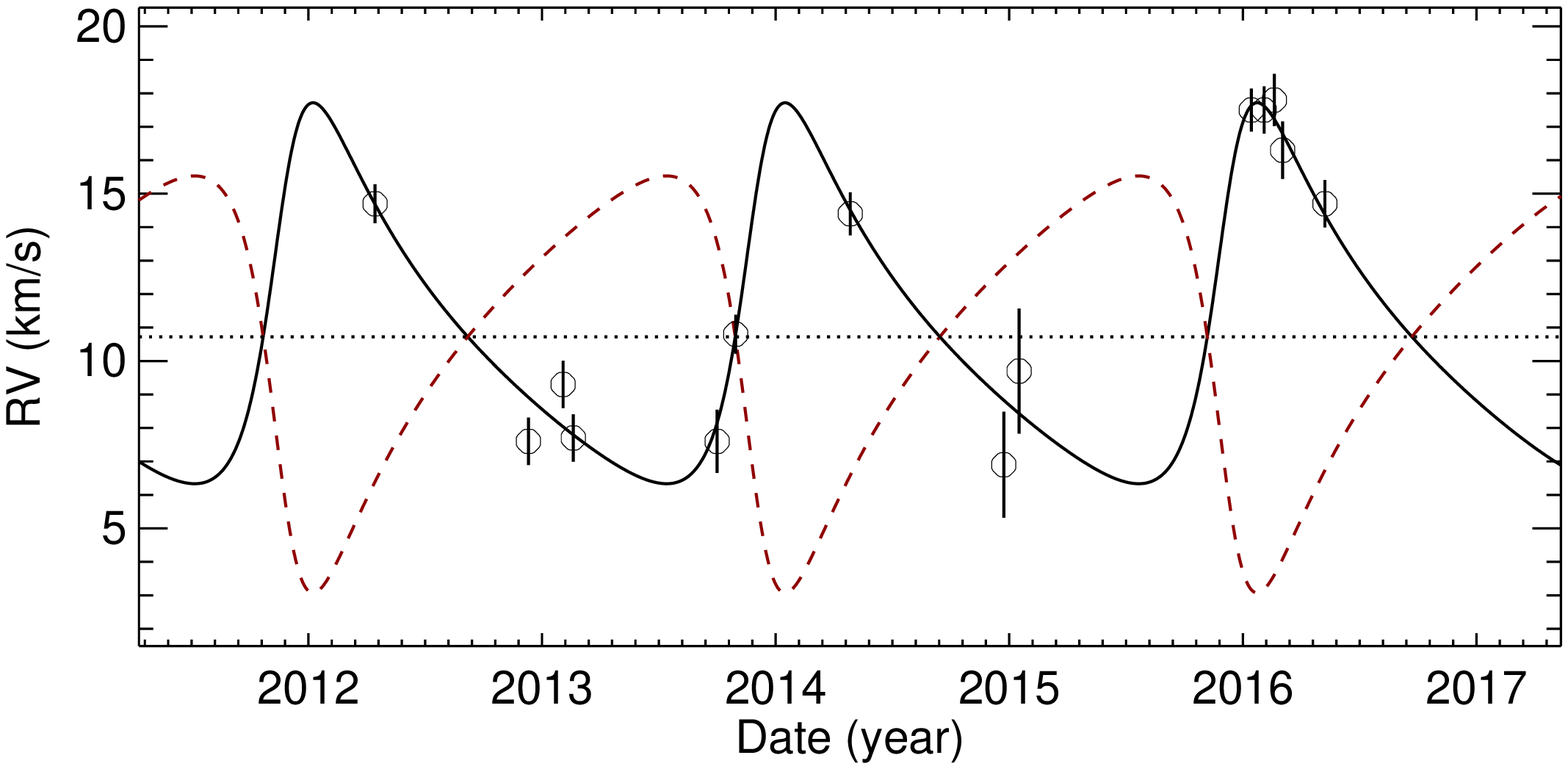} \\
\includegraphics[height=7cm]{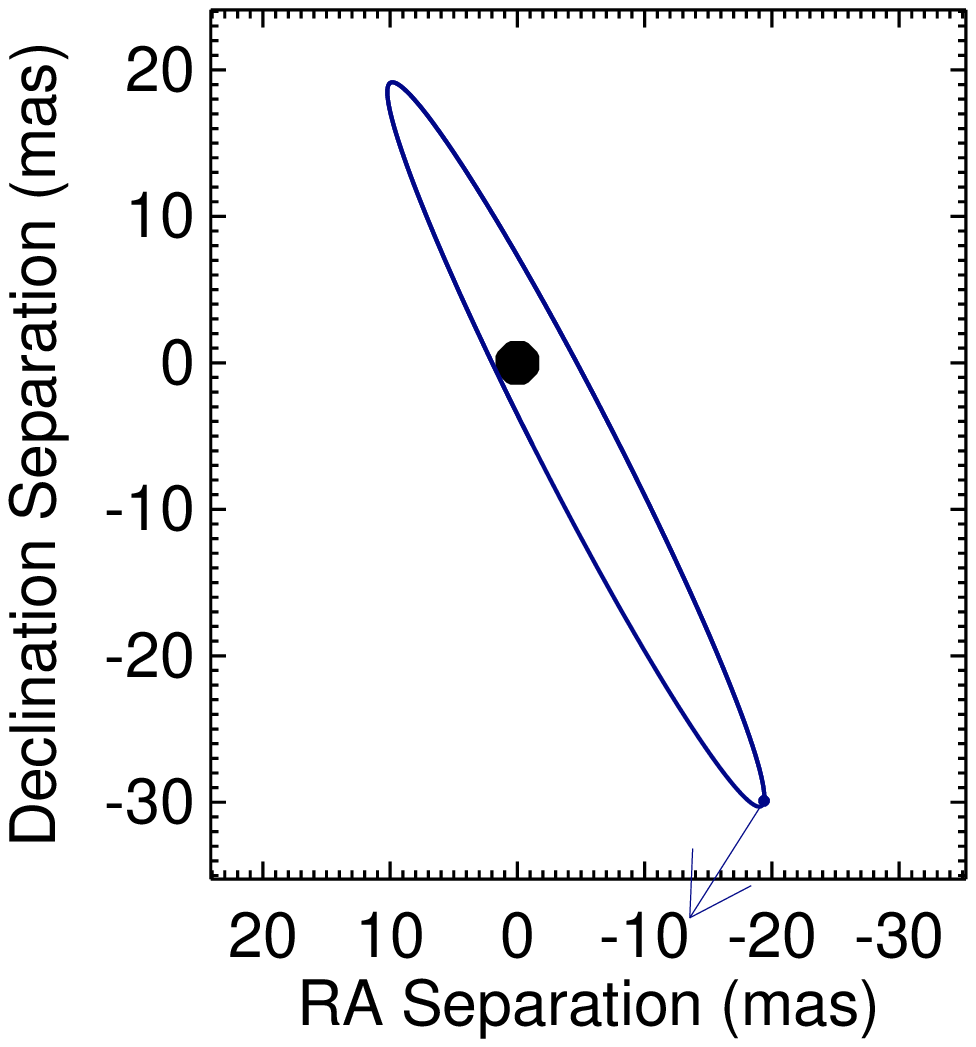} 
\includegraphics[height=7cm]{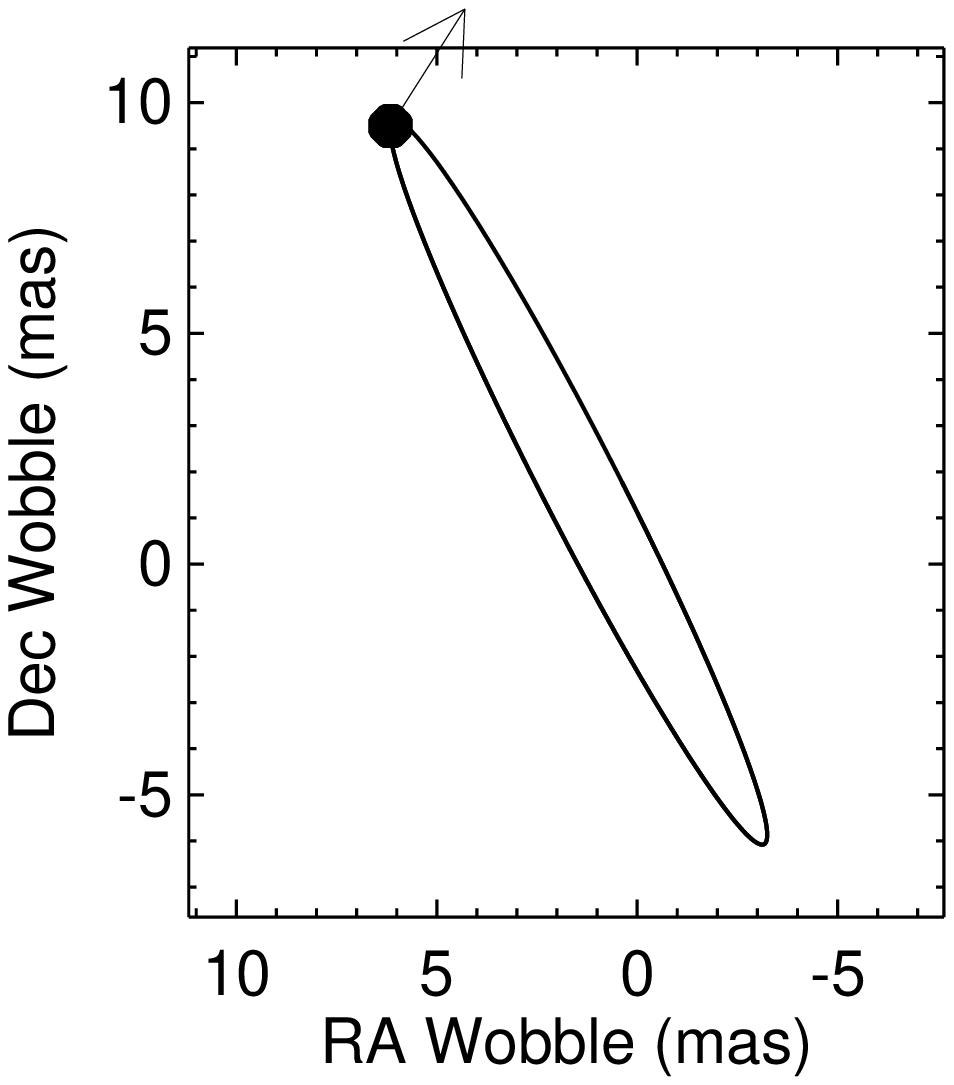} \\
\caption{Same as Figure~\ref{fig:orbit} but based on fits constrained by evolutionary models.
\label{fig:orbit_evol}}
\end{figure*}

\begin{figure}
\epsscale{1.0}
\centering
\plotone{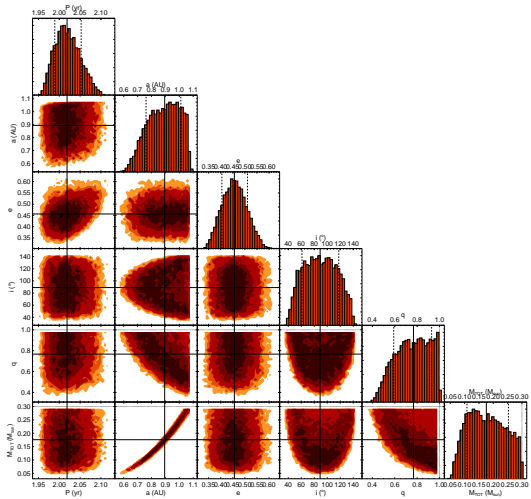}
\caption{Parameter distributions and correlations (triangle plot) for period ($P$), semi-major axis ($a$), eccentricity ($e$), 
inclination ($i$), mass ratio ($q$) and total system mass ($M_{tot}$) based on our MCMC orbital fit of the primary radial velocity data without constraints imposed by the evolutionay models. 
The fits assume weak constraints on period (0.2~yr $\leq P \leq$ 30~yr), eccentricity ($e \leq 0.95$) and total mass (M$_{tot}$ $\leq$ 0.3~{\msun}; dotted lines in histograms).
Contour plots show two-dimensional frequency distributions for parameter pairs, highlighting correlations (e.g., $a$ and M$_{tot}$, $q$ and M$_{tot}$).
Normalized histograms at the ends of rows are marginalized over all other parameters.
Median values are indicated by solid lines in all panels; 16\% and 84\% quantiles are indicated by dashed lines in the histograms. These values are listed in  Table~\ref{tab:orbit_parameters}.
\label{fig:orbit_parameters}}
\end{figure}

\begin{figure}
\epsscale{1.0}
\centering
\plotone{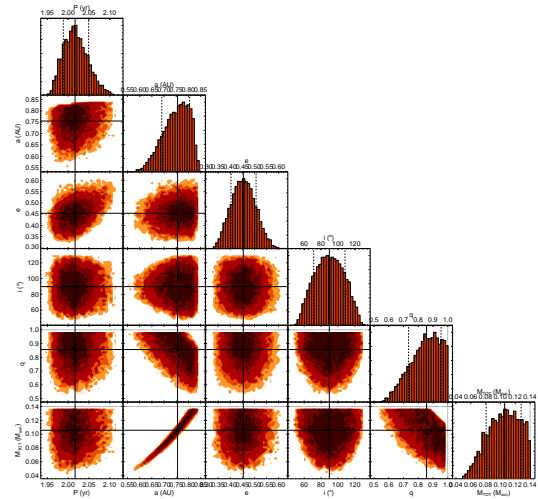}
\caption{Same as Figure~\ref{fig:orbit_parameters} but with orbit models constrained by 
evolutionary models.
\label{fig:orbit_parameters_evol}}
\end{figure}

Figures~\ref{fig:orbit} and~\ref{fig:orbit_evol} show the best-fit orbits from both analyses, while Figures~\ref{fig:orbit_parameters} and~\ref{fig:orbit_parameters_evol} display the distributions and correlations for $P$, $a$, $e$, $i$, $q$ and M$_{tot}$
from the MCMC chains.
Table~\ref{tab:orbit_parameters} lists the best-fit and mean orbital parameters and inferred component properties. The $\chi^2$ values for the best-fit solutions in both the constrained and unconstrained fits indicate convergence, and both analyses produce nearly identical values for the period ({2.02$\pm$~yr}), eccentricity (0.46$\pm$0.05), inclination\footnote{In this analysis, orbital inclinations $<$90$\degr$ correspond to clockwise orbital motion, $>$90$\degr$ to counterclockwise motion, as projected on the sky.} (90$\degr\pm19\degr$) and center-of-mass radial velocity ({10.8$\pm$0.3~{\kms}}). The remaining orbital values are also statistically consistent between the analyses. We verified that the period, an integer multiple of a year, was not the result of phased sampling; the opposing phase of our 2013 February and 2016 February measurements assures this. 

\begin{deluxetable*}{lcccc}
\tablecaption{Parameters from Orbital Analysis \label{tab:orbit_parameters}}
\tabletypesize{\small}
\tablewidth{0pt}
\tablehead{
 & \multicolumn{2}{c}{Without Evolutionary} & \multicolumn{2}{c}{With Evolutionary} \\
 & \multicolumn{2}{c}{Model Constraints} & \multicolumn{2}{c}{Model Constraints} \\
 \cline{1-5}
\colhead{Parameter} &
\colhead{Best-fit} &
\colhead{Median} &
\colhead{Best-fit} &
\colhead{Median}}
\startdata
\multicolumn{5}{c}{Modeled Parameters} \\
\hline
Best $\chi^2$ (DOF) & 10.3 (7) & \nodata & 10.1 (7) & \nodata\\
$P$\tablenotemark{a} (yr) & 2.02 &  2.02$^{+0.03}_{-0.03}$ & 2.02 &  2.02$^{+0.03}_{-0.03}$ \\
$a$ (AU) & 0.85  & 0.89$^{+0.12}_{-0.13}$ & 0.72  & 0.76$^{+0.05}_{-0.06}$ \\
$e$\tablenotemark{a} & 0.45 &  0.46$^{+0.05}_{-0.05}$ & 0.45 &  0.46$^{+0.05}_{-0.05}$ \\
$i$ ($\degr$) & 125 & 89$^{+29}_{-28}$  & 96 & 90$^{+19}_{-19}$ \\
$\omega$ ($\degr$) & 301 &  304$^{+16}_{-15}$ & 300 &  308$^{+15}_{-14}$ \\
$M_0$ ($\degr$) & 68 & 66$^{+13}_{-14}$  & 70 & 63$^{+13}_{-14}$ \\
$q$  & 0.93 &  0.77$^{+0.16}_{-0.18}$   & 0.91 &  0.86$^{+0.10}_{-0.12}$ \\
$V_{COM}$ ({\kms})  & 10.8 &  10.7$^{+0.3}_{-0.3}$   & 10.7 &  10.8$^{+0.3}_{-0.3}$ \\
\hline
\multicolumn{5}{c}{Inferred Parameters} \\
\hline
$M_{tot}$ ({\msun}) & 0.15 &  0.18$^{+0.08}_{-0.07}$   & 0.09 &  0.11$^{+0.02}_{-0.02}$  \\
$M_{1}$ ({\msun}) & 0.079 &  0.10$^{+0.05}_{-0.04}$  & 0.048 &  0.057$^{+0.016}_{-0.014}$ \\
$M_{2}$ ({\msun}) & 0.074 &  0.07$^{+0.03}_{-0.02}$  & 0.044 &  0.048$^{+0.008}_{-0.010}$ \\
$K_1$ ({\kms}) & 5.6 &  5.6$^{+0.6}_{-0.5}$  & 5.7 &  5.4$^{+0.4}_{-0.4}$  \\
$K_2$ ({\kms}) & 6.0  &  7.4$^{+2.1}_{-1.4}$   & 6.2  &  6.4$^{+1.0}_{-0.8}$  \\
Minimum Age (Gyr) &  \nodata & \nodata   & 4.2 & 4.0$^{+1.9}_{-1.2}$  \\
Minimum Inclination ($\degr$) & \nodata & \nodata & 64 & 63$^{+10}_{-8}$  \\
\enddata
\tablenotetext{a}{Parameter was constrained to a limited value range in MCMC analysis.}
\end{deluxetable*}

While the {two analyses yield equivalent orbital parameters}, they make notably different predictions for the inferred component properties of this system. The unconstrained fit favors larger values for the component and total system masses, predicting in particular a likely stellar mass for {\namesh}A and a mass at the hydrogen-burning limit for {\namesh}B.
In contrast, the constrained fit is limited to a total mass of 0.14~{\msun}, and {predicts masses for both components that are likely to be below} the hydrogen burning limit. These distinctions are discussed in Section~5. 

Returning to the orbital parameters, one striking feature of the fits is that an edge-on orbital inclination is favored, albeit with large uncertainties. This orientation is necessary for the constrained fit to reproduce the large radial velocity amplitude of the primary given the (model predicted) substellar masses of the components.  Requiring that the observed system mass function not exceed the maximum limit inferred from the evolutionary models, we determine a minimum system inclination of 63$\degr^{+10\degr}_{-8\degr}$.  This analysis also predicts a minimum age for the {\namesh} system of {4.0$^{+1.9}_{-1.2}$~Gyr}, again necessary to have a large enough secondary mass to reproduce the observed reflex motion of the primary.  We reiterate that these values are model-dependent, and also dependent on correct estimation of the component effective temperatures. Large inclinations and system ages were previously obtained in analyses of the spectral binaries 2MASS~J03202839$-$446358AB ($i \gtrsim 53\degr$, $\tau \gtrsim$ 2~Gyr; \citealt{2009AJ....137.4621B}) and SDSS~J000649.16$-$085246.3 ($i \gtrsim 61\degr$, $\tau \gtrsim$ 4~Gyr; \citealt{2012ApJ...757..110B}). The similarity in these results is likely a selection effect. Given the sensitivity limits to radial velocity variations for these low-mass and low-luminosity sources, older, more edge-on systems are more detectable as variables than younger and/or lower inclination systems.

\begin{figure}
\epsscale{0.8}
\centering
\plotone{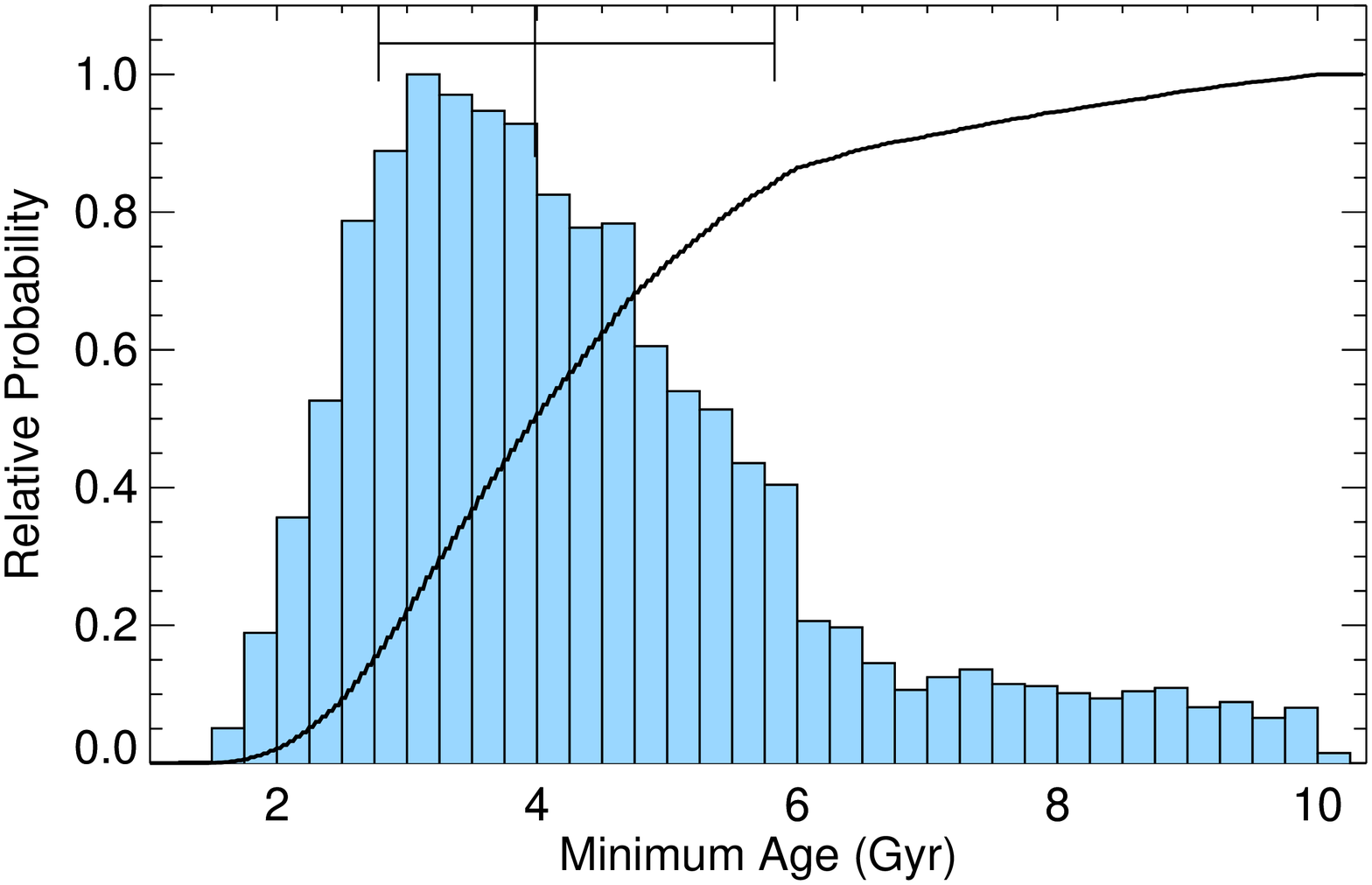} \\
\plotone{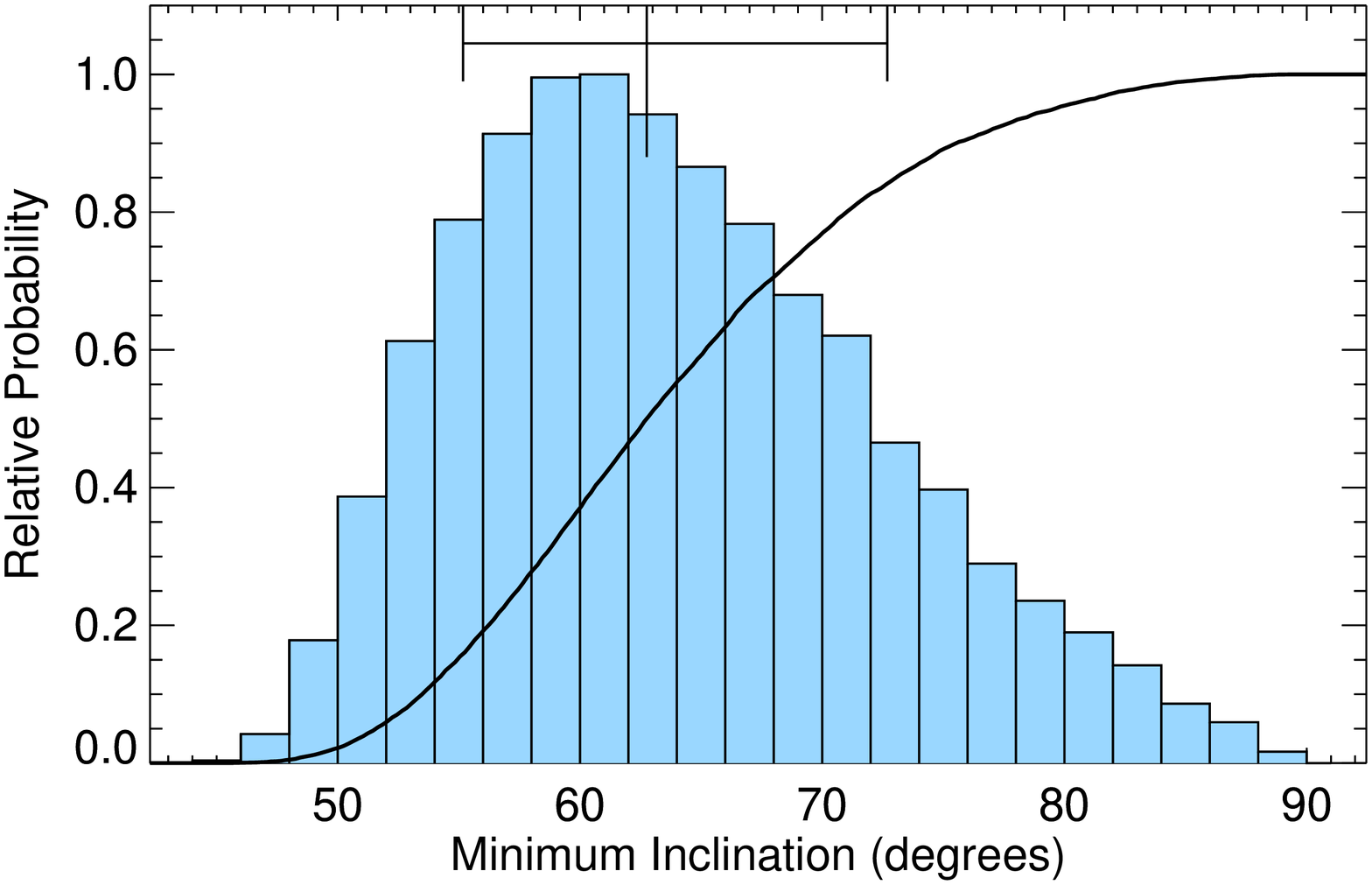}
\caption{Constraints on the minimum age (top panel) and minimum orbital plane inclination (bottom panel)
of the {\namesh}AB system, assuming limits imposed by the evolutionary models. The histograms show the distributions of these values for all viable orbits in the MCMC chain; the error bars at top indicates the median and 16\% and 84\% quantiles of the distributions (listed in Table~\ref{tab:orbit_parameters}). Red lines trace the cumulative distributions.
\label{fig:minageinc}}
\end{figure}

Finally, we note that the predicted astrometric orbits for the best-fit cases are consistent with prior measurements of the system. The separation between primary and secondary as projected on the sky does not exceed 50~mas, well within the resolution limits of previously published LGSAO observations. {The astrometric} wobble of the primary, assuming a relative magnitude of $\Delta{J}$ = 1.8~mag based on the spectral template analysis above, has a maximum amplitude of roughly 15~mas, consistent with the astrometric residuals reported by \citet{2012ApJS..201...19D}. 
  
\section{Discussion\label{sec:discuss}}

The combined detection of astrometric and radial velocity variability unambiguously confirms {\namesh} as a VLM binary system. Our measurements of the primary radial motion yield stringent constraints on the orbital period ({1.4\%}), system velocity (3\%), {eccentricity (5\%),} and semi-major axis (8\%); and {a reasonable constraint on the system mass ratio (12\%) which is coupled to other parameters ($a$ and M$_{tot}$) and partly dependent on the atmosphere models.}  The inferred system and component masses are much more weakly constrained (20\%--30\%), strongly correlated with other parameters (e.g., $q$), and highly sensitive to inputs from the evolutionary models. As such, our orbit parameter determinations are insufficient to directly test the models. 

There are additional observables that could be brought to bear on this problem, however. The inferred primary mass is either below or above the lithium-burning minimum mass limit of 0.060~{\msun} \citep{1997ApJ...482..442B,1998ApJ...497..253U}, depending on whether the orbit fits are constained by evolutionary models or not. The Li~I absorption line at 6708~{\AA} is readily detectable in the optical spectra of early- and mid-type L dwarfs; however, previously reported spectral data of {\namesh} do not have sufficient S/N to assess the presence of this feature. New observations to apply the ``lithium test'' on this system \citep{1991A&A...249..149M,1994ApJ...436..262M,2005ApJ...634..616L} may considerably constrain the allowed parameter space for its orbit, or indicate disagreement between the evolutionary models and orbital parameters.
A more direct measure of the mass ratio could also be made from the reflex motion of the secondary, whose signal is buried within the combined light spectrum of the system. Analysis methods such as TODCOR \citep{1994Ap&SS.212..349M} could be employed to extract this signal in spectral regions where the secondary contributes a greater fraction of the total flux, such as the 1.25--1.30~{\micron} $J$-band and 1.55--1.60~{\micron} $H$-band psuedo-continuum peaks of T dwarfs. These observations are currently proposed and will be examined in a future study.

Despite the accuracy obtained for the orbital elements $P$, $a$ and $e$, further observations to more tightly constrain the orbit geometry are warranted. Of particular interest is inclination, as the rapid rotation of {\namesh}A inferred from these measurements makes this an ideal system to explore spin-orbit alignment in VLM star-brown dwarf multiples. To date, only a single VLM stellar pair, the L0+L1.5 2MASSW~J0746425+200032AB \citep{2000AJ....119..369R,2001AJ....121..489R,2003AJ....126.1526B,2004A&A...423..341B} has been tested and confirmed to be in alignment to within 10$\degr$ \citep{2013A&A...554A.113H}. Improving the constraint on the inclination of {\namesh}AB to within this limit, and measuring a robust rotation period through photometric variability,\footnote{\citet{2013AJ....145...71K} reported no variability in two monitoring epochs of 2~hr each to a limiting amplitude of 3\% in both $J$- and $K$-band observations; however,  this limit is comparable to the amplitudes of known VLM variables \citep{2012ApJ...750..105R}.} would permit a similar test of alignment based on an assumed radius, or a radius determination for {\namesh}A assuming alignment. 
Combining our radial velocity measurements with prior or concurrent measurements of astrometric variability should in principle improve orbital parameters, as well as yield a measure of the longitude of ascending node which is unconstrained in these data.  
While a direct view of the orbit has so far proven too challenging for LGSAO direct imaging, sparse-aperture mask imaging with AO \citep{2006SPIE.6272E.103T} may be a useful alternative approach. Prior work has demonstrated that companions with contrast ratios of 
$\Delta{m} \gtrsim$ 3, appropriate for $K$-band imaging of {\namesh}AB, can be resolved for separations $\gtrsim$20~mas with Keck NIRC2 LGSAO \citep{2008ApJ...679..762K,2008ApJ...681..579B}. This is sufficient to resolve the system at apoapsis, and would again aid in constraining the overall orientation of the orbit.

The semi-major axis of this system falls well below the peak of the separation distribution of the current sample of known VLM multiples, $\sim$4~AU \citep{2007ApJ...668..492A,2014ApJ...794..143B,2015AJ....150..163B}. Since this sample is dominated by sources uncovered through resolved imaging, our result is not particularly surprising. However, it does add to growing evidence that tight separations are common among confirmed VLM spectral binary systems, whose identification is independent of separation up to $\sim$500~mas (10~AU at 20~pc). These results suggest that many other VLM systems without the necessary spectral composition to be detected as spectral binaries may be currently overlooked.  Ongoing radial velocity monitoring, astrometric monitoring and high-resolution imaging of spectral binary candidates will provide a more robust assessment of the close-separation binary fraction, and a pathway toward accurate determination of the overall binary fraction of the coolest stars and brown dwarfs.

\acknowledgements
The authors thank Joel Aycock, Scott Dahm, Randy Campbell, Greg Doppman, Heather Hershey, Carolyn Jordan, Marc Kassis, Jim Lyke, Gary Punawai, Julie Rivera, Terry Stickel, Hien Tran, and Cynthia Wilburn at Keck Observatory, and Christine Nichols and Melisa Tallis at UCSD,
for their assistance with the NIRSPEC observations.
A.J.B. acknowledges funding support from the National Science Foundation under award No.\ AST-1517177. 
The material is based upon work supported by the National Aeronautics and Space Administration under Grant No. NNX15AI75G.
This research has made use of the SIMBAD database,
operated at CDS, Strasbourg, France;
NASA's Astrophysics Data System Bibliographic Services;
the M, L, T, and Y dwarf compendium housed at DwarfArchives.org;
and the SpeX Prism Libraries at \url{http://www.browndwarfs.org/spexprism}.
{We thank our anonymous referee for her/his/their prompt and helpful review of the original manuscript.}
The authors recognize and acknowledge the very significant cultural role and reverence that the summit of Mauna Kea has always had within the indigenous Hawaiian community.  We are most fortunate and grateful to have the opportunity to conduct observations from this mountain.

\appendix

\section{Forward Modeling of the NIRSPEC Spectra}

To accurately determine radial and rotational velocities from the NIRSPEC data, we adapted the forward-modeling procedure described in \citet{2015AJ....149..104B}, which is in turn based on the method described in \citet{2010ApJ...723..684B}.  Data were initially reduced and rectified using a modified version of the REDSPEC package, and source and A0 standard spectra in order~33 (2.29--2.33~$\micron$) optimally extracted, scaled and co-added with uncertainty weighting. Spectral uncertainties ($\sigma$) were determined from image variance, a combination of Poisson shot noise, read noise (50~e$^-$; \citealt{2010ApJ...723..684B}) and variance between the individual extractions.

These ``raw'' spectra are a function of pixel position and include telluric absorption and residual pixel sensitivity variations. Rather than calibrate these effects, we followed an iterative forward-modeling approach using a multi-threaded MCMC method with a Metropolis-Hasting algorithm \citep{1953JChPh..21.1087M,HASTINGS01041970}.  The extracted data were modeled as
\begin{equation}
D[p(\lambda)] = C[p(\lambda)] \times \left[ \left( M[p(\lambda[1+\frac{RV}{c}])] \ast \kappa_R(V_{rot}\sin{i}) \right) \times T[p(\lambda)]^{\alpha} \right] \ast \kappa_G(\Delta{v}_{inst}).
\end{equation}
Here, $p(\lambda)$ is the wavelength-to-pixel translation, which was modeled as a 
second-order polynomial; $C[p]$ is a continuum correction, also modeled as a second-order polynomial; 
$M[p]$ is a solar-metallicity BT-Settl atmosphere model \citep{2011ASPC..448...91A} parameterized by {\teff} and {\logg}, used to represent the spectrum of {\namesh}; the model spectrum is wavelength-shifted by the radial velocity {\rv};
$T[p]$ is the telluric transmission spectrum from the Solar atlas of \citet{1991aass.book.....L};
$\alpha$ is the telluric transmission scaling factor; and
$\kappa_R$ and $\kappa_G$ are the rotational and instrumental broadening profiles convolved ($\ast$) with the model spectrum, which are parameterized by the projected rotational velocity {\vsini} and a Gaussian with velocity width $\Delta{v}_{inst}$, respectively.

\begin{figure}
\epsscale{0.8}
\centering
\plotone{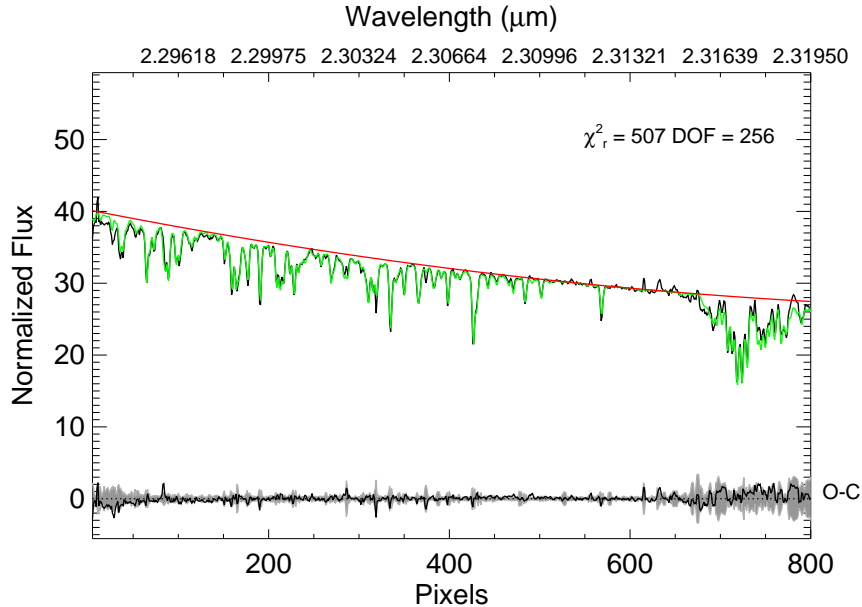}
\caption{Fit to the spectrum of the A0~V telluric calibrator HD~71906 observed on 2016 February 16 (UT). The extracted spectrum is shown in black; the (featureless) A0~V continuum, modeled as a second-order polynomial, is shown in red;
and the full model, including scaled telluric absorption, is shown in green. Residuals are plotted as the grey
line around zero, and is dominated by uncorrected fringing. Pixel scale is listed along the bottom while wavelength scale is listed along the top.
\label{fig:calfit}}
\end{figure}

The full model contains 12 parameters, but not all were fit simultaneously.
We first determined the wavelength-to-pixel translation and instrumental broadening using our calibration observations. This mapping was first estimated by comparing the arc lamp spectrum to line air wavelengths  as compiled by the National Institute of Standards and Technology (NIST Atomic Line Database; \citealt{NIST_ASD}). We then fit the telluric absorption spectrum of the otherwise featureless A0~V star ($M[p]$ = 1), iteratively fitting residuals in cross-correlations between the Solar telluric atlas and the extracted telluric spectrum over narrow (30~pixel = 0.001~$\micron$) spectral regions to converge on the wavelength solution. Typical residuals were (0.7-1.0)$\times$10$^{-6}$~{\micron} ($\sim$0.1~{\kms}). The instrumental broadening was also determined at this step to be in the range 4.6--5.0 pixels ($\Delta{v}_{inst}$ = 19--21~{\kms}).  An example fit from data on 2016 February 16 (UT) is shown in Figure~\ref{fig:calfit}.

The spectrum of {\namesh}AB was fit in three separate passes. First, fixing the instrumental broadening and first- and second-order coefficients for the wavelength solution, the four parameters {\rv}, {\vsini}, $\alpha$ and the zeroth-order coefficient in the wavelength-to-pixel translation, as well as the three coefficients for the continuum correction, were determined by MCMC analysis using a {\teff} = 1700~K, {\logg} = 5.0 cgs model and initial estimates {\rv} = 0~{\kms} and {\vsini} = 30~{\kms}. A single MCMC chain of length 8,000 steps was used (2,000 per parameter), with parameters sequentially updated (Gibbs sampling) by drawing offsets from normal distributions with pre-determined widths (generally $\sim$10 times larger than the final uncertainties). A chi-square statistic was used to compare data ($d[p]$) to model ($D[p]$)
\begin{equation}
\chi^2 = \sum\frac{(d[p]-D[p])^2}{\sigma[p]^2}
\end{equation}
A new parameter set $\vec\theta(i) \rightarrow \vec\theta(i+1)$ was adopted if the acceptance condition $U(0,1) \leq e^{-0.5(\chi^2(i+1)-\chi^2(i))}$ was satisfied, where $U(0,1)$ is a random draw from a uniform distribution between 0 and 1.
The effective degrees of freedom of this fit was estimated as (number of data pixels)/3 - (number of fit parameters) $\approx$ 250.  The scale factor of 3 pixels roughly accounts for correlated data due to instrumental line broadening. 
Note that the coefficients for the continuum correction function were not iterated in this manner, but determined 
instead by fitting a second-order polynomial to the ratio of model and observed spectrum at each step. 
After this initial chain, these parameters were fixed and the data then compared to a suite of BT-Settl models spanning {\teff} = 1500--2500~K in steps of 100~K and {\logg} = 4.0 to 5.5 (cgs) in steps of 0.5~dex, again using the $\chi^2$ statistic and fitting the continuum separately.  The best-fit atmosphere model from this analysis, typically {\teff} =1900-2100~K and {\logg} = 5.0--5.5, was then used as the starting point of a multi-threaded MCMC analysis for which 9 parameters ({\rv}, {\vsini}, $\alpha$, {\teff}, {\logg}, the zeroth-order wavelength coefficient, and the three continuum coefficients) were fit iteratively. {We used an implementation of the \citet{goodman2010} affine-invariant MCMC ensemble with $N_C$ = 10 independent chains, each with initial conditions drawn from uniform distributions centered on the best-fit values from the first two fitting passes and widths at least three times the standard deviation of these passes.} Models with intermediate values of {\teff} and {\logg} were linearly interpolated in logarithmic flux between the model grid nodes. The chains were evolved for $N_S$ = 12,000 steps following the same Metropolis-Hastings algorithm as above.

\begin{figure}
\epsscale{1.1}
\centering
\plottwo{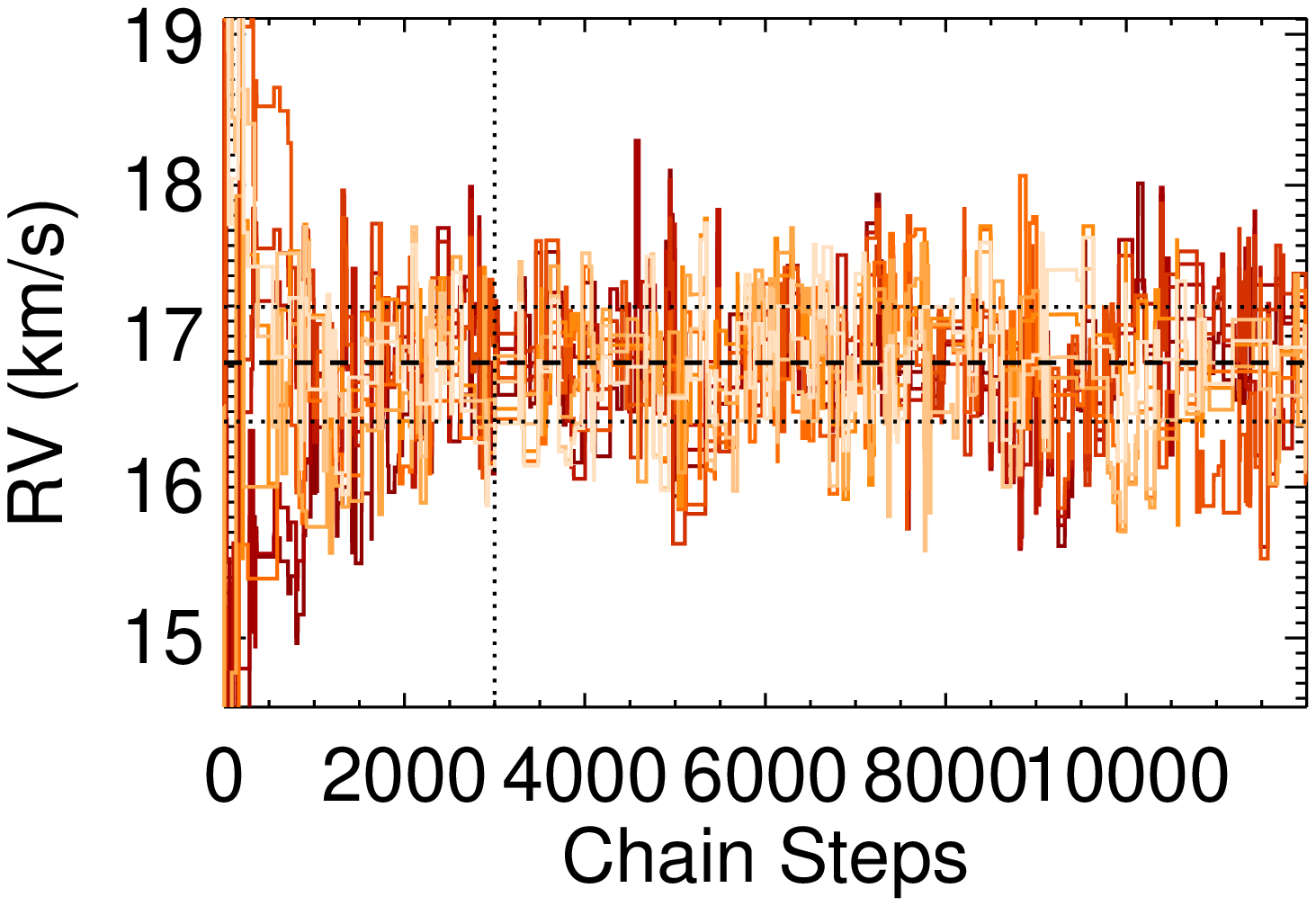}{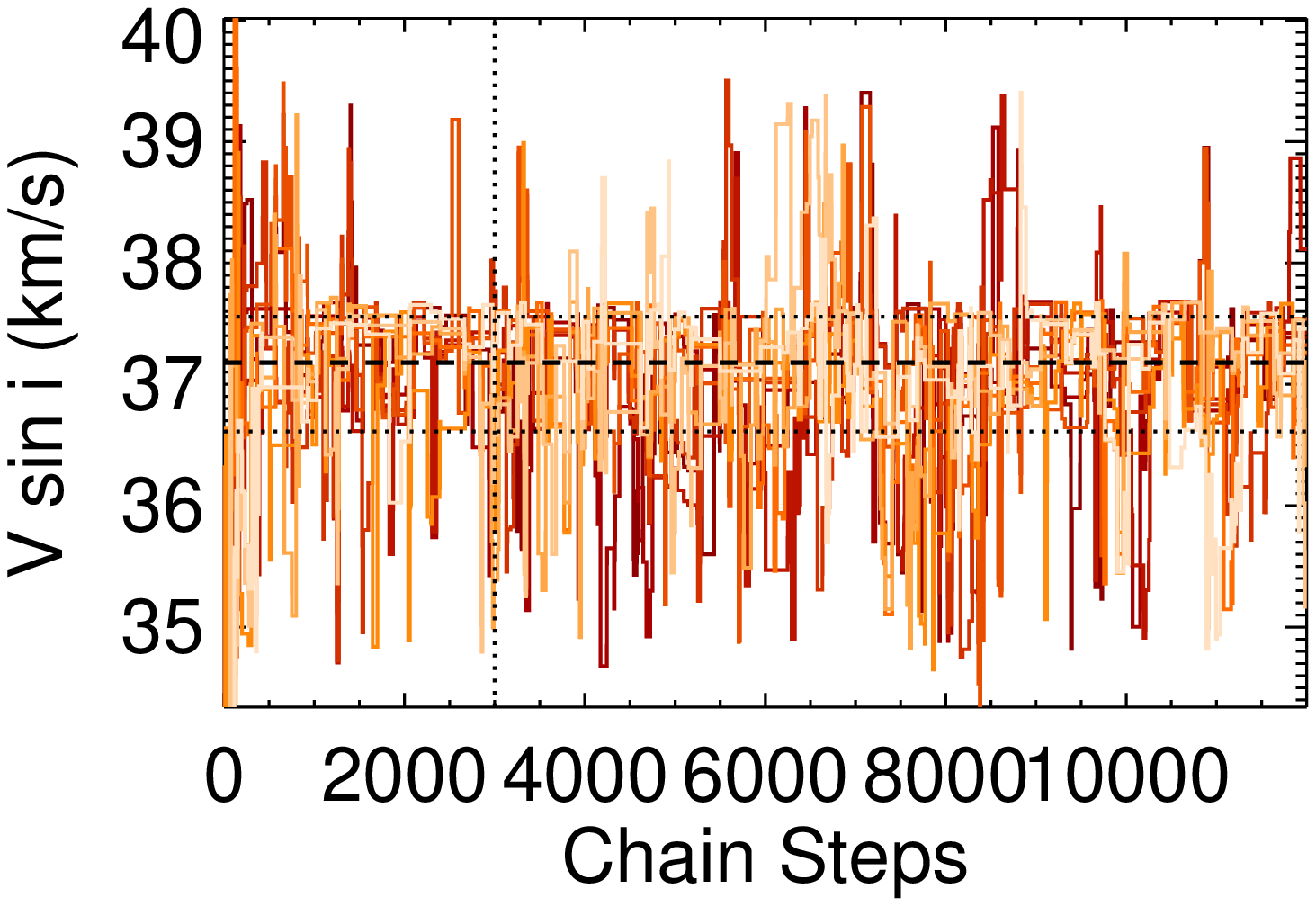} \\
\plottwo{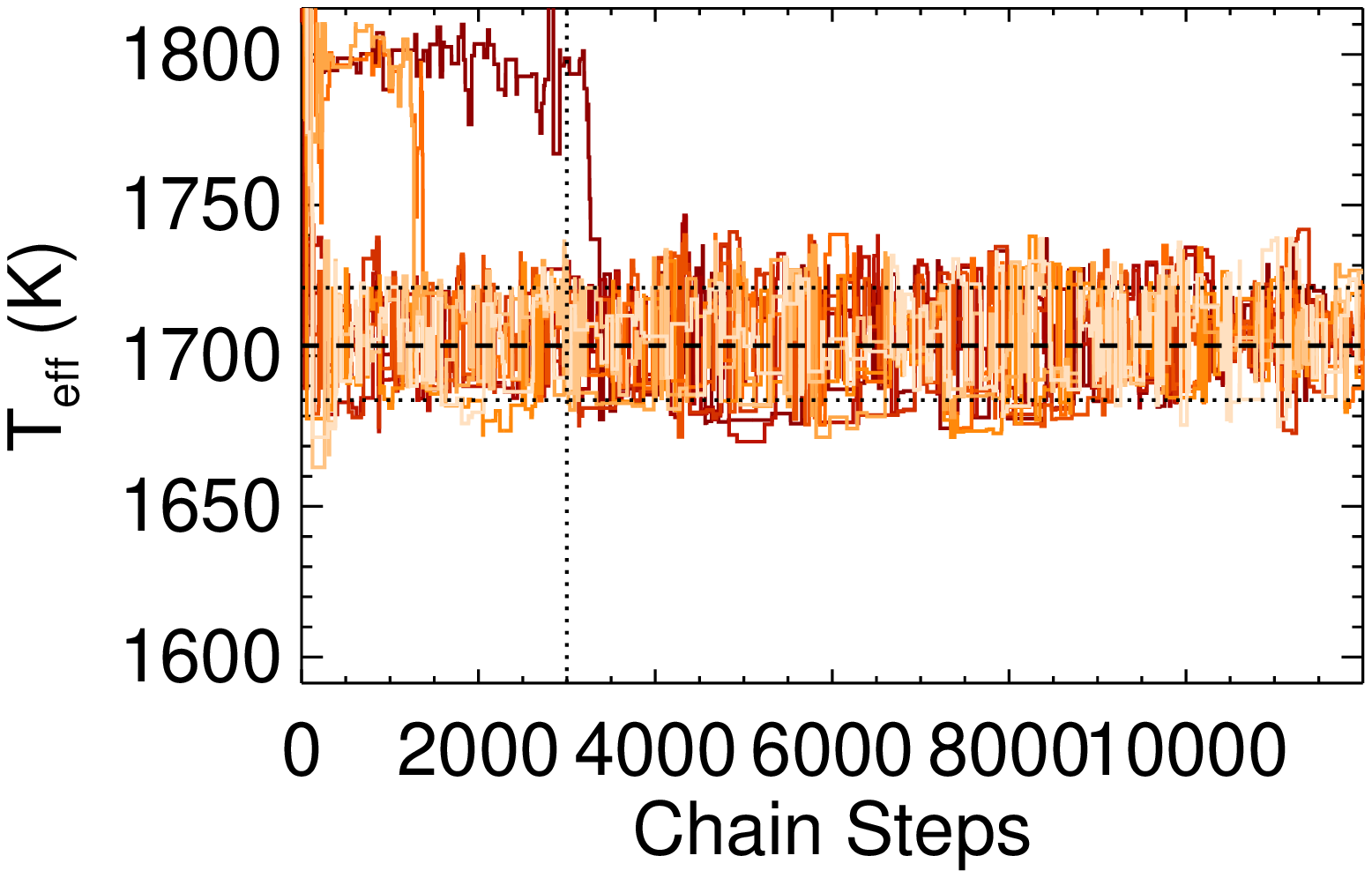}{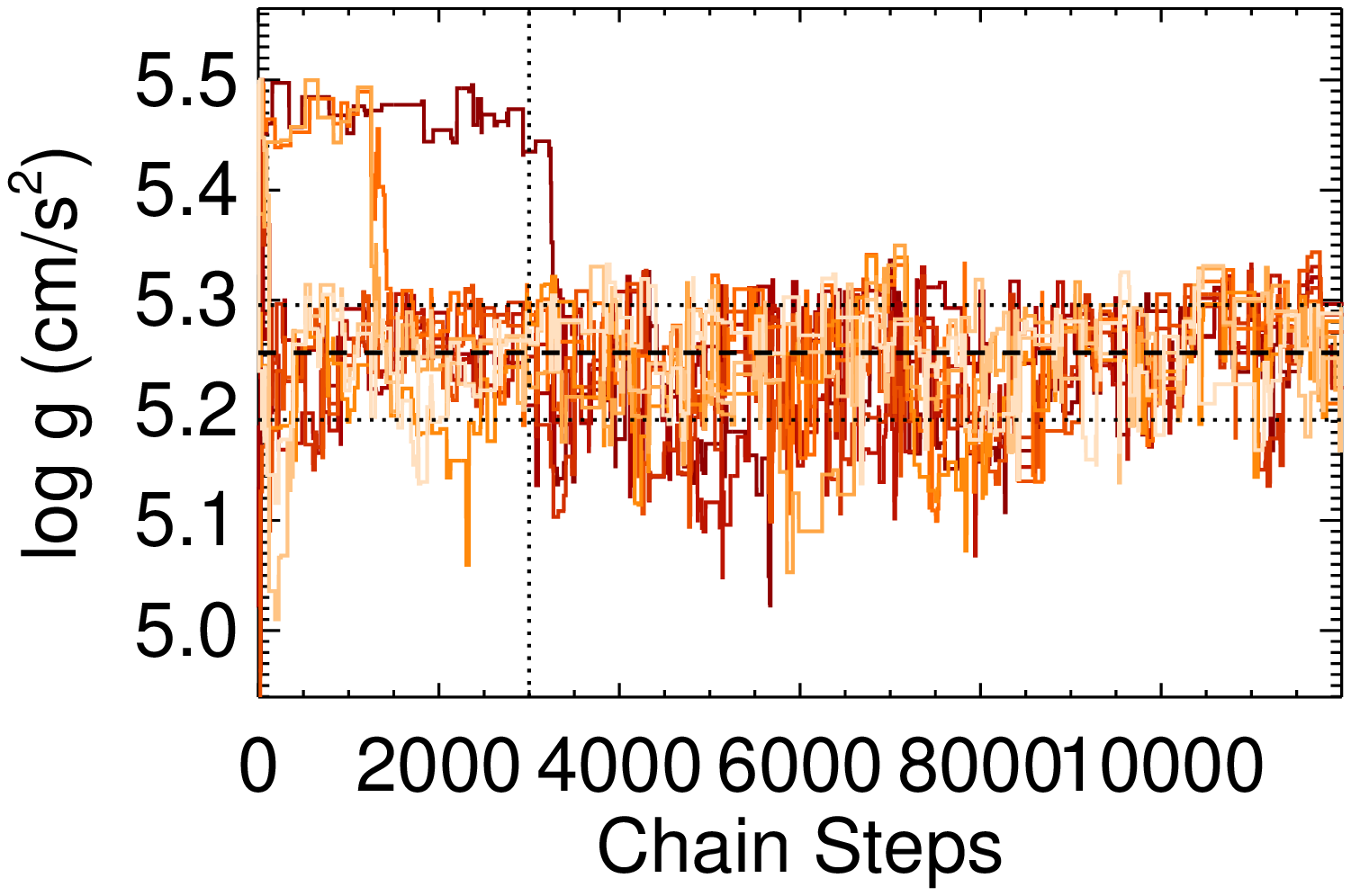}
\caption{MCMC chains for fit parameters {\rv}, {\vsini}, {\teff} and {\logg} for data taken on 2016 February 16 (UT). The best-fit spectrum is shown in Figure~\ref{fig:nirspec}.  Chain values to the left of the dotted lines were not included in the parameter distributions and estimates.
\label{fig:mcmc_chains}}
\end{figure}

\begin{figure}
\epsscale{1.0}
\centering
\plotone{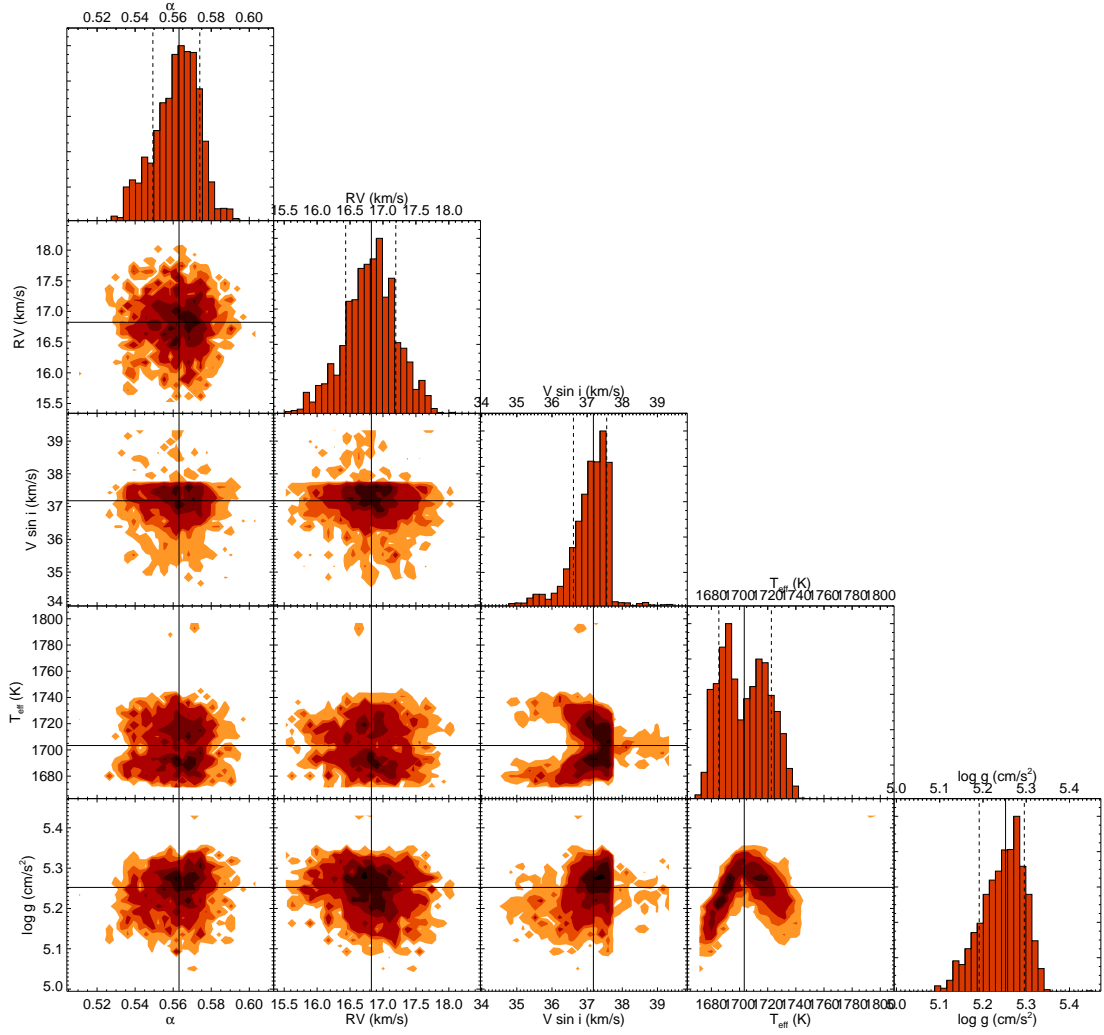}
\caption{Parameter distributions and correlations for fit parameters {\rv}, {\vsini}, {\teff} and {\logg} for data taken on 2016 February 16 (UT). 
Contour plots show two-dimensional frequency distributions for parameter pairs, highlighting correlations.
Normalized histograms at the ends of rows are marginalized over all other parameters.
Median values are indicated by solid lines in all panels; 16\% and 84\% quantiles are indicated by dashed lines in the histograms.
\label{fig:mcmc_triangle}}
\end{figure}
\clearpage

Figure~\ref{fig:mcmc_chains} shows the chain evolution for the parameters {\rv}, {\vsini}, {\teff} and {\logg} for our fit to the 2016 February 16 (UT) data, while Figure~\ref{fig:mcmc_triangle} compares the resulting distributions of these parameters based on the last 75\% of all chains.
The chains for {\rv} and {\vsini} converge quickly to common values, and 
convergence was quantified for all parameters by computing the \citet{doi:10.1214/ss/1177011136} scale reduction factor for each parameter $\theta_j$,
\begin{equation}
\hat{R}_j = \frac{N_S-1}{N_S} + \frac{N_C+1}{N_SN_C}\frac{B_j}{W_j}\frac{DOF}{DOF-2}
\end{equation}
where 
\begin{equation}
W_j = \frac{1}{N_C}\sum_{i=1}^{N_C}\left(\theta_{ij}-\bar\theta_{ij}\right)^2
\end{equation}
\begin{equation}
B_j = \frac{N_S}{N_C-1}\sum_{i=1}^{N_C}\left(\bar\theta_{ij}-\bar\theta_j\right)^2
\end{equation}
are the average within-chain variances and variance in the between-chain means, respectively.\footnote{$\bar\theta_{ij}$ is the average of parameter $\theta_j$ for chain $i$; $\bar\theta_{j}$is the average of parameter $\theta_j$ across all chains.}. For {\rv}, {\vsini}, and the other fitting parameters, we confirmed that ${\hat{R}}$ $<$ 1, indicating that the well-sampled prior distribution converged to a common posterior distribution. The model parameters {\teff} and {\logg}, on the other hand, converged more slowly and, in some cases, to distinct, discrete values for different chains indicating that these parameters are not properly converged. We find minimal correlation between these model parameters and {\rv} and {\vsini} (Figure~\ref{fig:mcmc_triangle}), so we did not attempt to address this issue, and defer discussion for a future study.
The best-fit and mean parameters and their uncertainties for all epochs are summarized in Table~\ref{tab:specfit_results}.

\begin{deluxetable*}{lccccccccc}
\tablecaption{Fit Parameters for All NIRSPEC Observations\label{tab:specfit_results}}
\tabletypesize{\scriptsize}
\tablewidth{0pt}
\tablehead{
& \multicolumn{3}{c}{Wavelength Solution} \\ \cline{2-4}
\colhead{UT Date} &
\colhead{c$_0$} &
\colhead{c$_1$} &
\colhead{c$_2$} &
\colhead{$\Delta{v}_{inst}$} &
\colhead{$\alpha$} &
\colhead{RV} &
\colhead{\vsini} &
\colhead{\teff} &
\colhead{\logg} \\
& 
\colhead{(pix)} &
\colhead{(pix~{\micron}$^{-1}$)} &
\colhead{(pix~{\micron}$^{-2}$)} &
\colhead{({\kms})} & &
\colhead{({\kms})} &
\colhead{({\kms})} &
\colhead{(K)} &
\colhead{(cgs)} \\
}
\startdata
2012 Apr 02 & 977.55 & 33528.4 & 104426 & 18.1 & 0.565 & 14.3 & 37.7 & 1712 & 4.9 \\
 & 977.52$\pm$0.05 & \nodata & \nodata & \nodata  & 0.558$\pm$0.012 & 14.7$\pm$0.3 & 38.0$\pm$0.4 & 2027$\pm$156 & 5.28$\pm$0.17 \\
2012 Nov 27 & 971.57 & 33448.6 & 103661 & 20.0 & 0.613 & 7.5 & 37.5 & 1690 & 5.3 \\
 & 971.59$\pm$0.06 & \nodata & \nodata & \nodata  & 0.614$\pm$0.014 & 7.4$\pm$0.5 & 37.2$\pm$0.7 & 1722$\pm$32 & 5.35$\pm$0.05 \\
2013 Jan 20 & 959.50 & 33399.4 & 103745 & 20.3 & 0.523 & 9.1 & 36.6 & 1698 & 5.1 \\
 & 959.48$\pm$0.08 & \nodata & \nodata & \nodata  & 0.523$\pm$0.012 & 9.1$\pm$0.5 & 36.1$\pm$0.8 & 1699$\pm$9 & 5.09$\pm$0.04 \\
2013 Feb 05 & 959.65 & 33414.9 & 103817 & 20.1 & 0.626 & 7.6 & 37.3 & 1902 & 4.6 \\
 & 959.66$\pm$0.06 & \nodata & \nodata & \nodata  & 0.625$\pm$0.014 & 7.6$\pm$0.4 & 37.0$\pm$1.2 & 1920$\pm$50 & 4.67$\pm$0.09 \\
2013 Sep 17 & 962.59 & 33413.4 & 103221 & 20.8 & 0.802 & 8.0 & 37.1 & 2026 & 5.3 \\
 & 962.59$\pm$0.08 & \nodata & \nodata & \nodata  & 0.80$\pm$0.02 & 7.8$\pm$0.7 & 37.3$\pm$1.8 & 2019$\pm$169 & 5.27$\pm$0.20 \\
2013 Oct 16 & 962.39 & 33405.0 & 103065 & 18.9 & 0.628 & 10.6 & 37.6 & 1689 & 5.1 \\
 & 962.41$\pm$0.04 & \nodata & \nodata & \nodata  & 0.626$\pm$0.010 & 10.5$\pm$0.3 & 37.2$\pm$0.5 & 1714$\pm$172 & 5.12$\pm$0.17 \\
2014 Apr 13 & 961.84 & 33425.0 & 103765 & 17.5 & 0.550 & 14.2 & 36.7 & 1799 & 5.5 \\
 & 961.84$\pm$0.03 & \nodata & \nodata & \nodata  & 0.552$\pm$0.010 & 14.4$\pm$0.4 & 35.9$\pm$0.9 & 2016$\pm$108 & 5.490$\pm$0.013 \\
2014 Dec 08 & 957.53 & 33343.8 & 102179 & 18.9 & 0.778 & 6.7 & 26.2 & 1715 & 5.3 \\
 & 957.56$\pm$0.07 & \nodata & \nodata & \nodata  & 0.78$\pm$0.02 & 6.6$\pm$0.5 & 26.7$\pm$0.9 & 1711$\pm$17 & 5.27$\pm$0.05 \\
2015 Jan 01 & 963.11 & 29015.8 & -41295 & 20.0 & 0.585 & 9.3 & 28.4 & 1727 & 5.1 \\
 & 963.05$\pm$0.09 & \nodata & \nodata & \nodata  & 0.59$\pm$0.03 & 9.7$\pm$0.8 & 28.4$\pm$1.7 & 1717$\pm$26 & 5.10$\pm$0.16 \\
2015 Dec 29 & 955.53 & 33406.1 & 104010 & 19.7 & 0.563 & 17.5 & 39.1 & 1700 & 5.2 \\
 & 955.54$\pm$0.05 & \nodata & \nodata & \nodata  & 0.560$\pm$0.011 & 17.5$\pm$0.4 & 39.0$\pm$0.7 & 2031$\pm$164 & 5.48$\pm$0.13 \\
2016 Jan 18 & 953.59 & 33347.4 & 102526 & 20.5 & 0.583 & 17.7 & 36.3 & 1700 & 5.3 \\
 & 953.63$\pm$0.06 & \nodata & \nodata & \nodata  & 0.587$\pm$0.017 & 17.4$\pm$0.4 & 36.1$\pm$0.9 & 1707$\pm$42 & 5.26$\pm$0.11 \\
2016 Feb 03 & 956.44 & 33354.6 & 102749 & 20.2 & 0.578 & 17.7 & 36.1 & 1694 & 5.2 \\
 & 956.48$\pm$0.08 & \nodata & \nodata & \nodata  & 0.57$\pm$0.02 & 17.7$\pm$0.5 & 35.4$\pm$1.0 & 1708$\pm$18 & 5.24$\pm$0.05 \\
2016 Feb 16 & 957.08 & 33390.0 & 102634 & 19.1 & 0.566 & 16.9 & 37.5 & 1691 & 5.2 \\
 & 957.08$\pm$0.05 & \nodata & \nodata & \nodata  & 0.563$\pm$0.012 & 16.8$\pm$0.4 & 37.2$\pm$0.5 & 1703$\pm$19 & 5.25$\pm$0.05 \\
2016 Apr 22 & 959.54 & 33405.9 & 103519 & 19.1 & 0.658 & 14.6 & 39.7 & 1702 & 5.2 \\
 & 959.52$\pm$0.04 & \nodata & \nodata & \nodata  & 0.656$\pm$0.012 & 14.7$\pm$0.5 & 39.0$\pm$0.8 & 1802$\pm$161 & 5.46$\pm$0.14 \\
\enddata
\tablecomments{These are the fit parameters emerging from the final multi-threaded MCMC fits for each observed spectrum, excluding the coefficients for the continuum correction which were determined dynamically. The first row for each date lists the best-fit (lowest $\chi^2$) parameters; the second row lists the means and standard deviations across all retained parameters in the MCMC chains. The coefficients for the wavelength-to-pixel conversion are defined as $p(\lambda) = \sum_{i=0}^2c_i(\lambda-\lambda_0)^i$, where $\lambda_0$ = 2.32428~{\micron}.  These coefficients, and the instrumental broadening, were not varied in the final MCMC fit.  None of the uncertainties listed for the mean values include systematic errors, which are estimated as 0.5~{\kms} for {\rv} and {\vsini}, 50~K for {\teff}, and 0.25~dex for {\logg}.}
\end{deluxetable*}

\section{Orbit Fitting Analysis}

The primary radial velocity orbit of {\namesh}AB was inferred using an adaptation of the MCMC analysis described in \citet{2012ApJ...757..110B,2015AJ....149..104B}. We examined a two-component orbit model with seven parameters,
\begin{equation}
\vec{\theta} =  \left(P,a,e,i,\omega,M_0,q,V_{COM}\right)
\end{equation}
where $P$ is the period of the orbit in years, $a$ is the semi-major axis in AU, $e$ is the eccentricity, $i$ is the inclination, $\omega$ is the argument of periastron, $M_0$ is the mean anomaly at epoch $\tau_0$ = 2012.253 (MJD\footnote{Modified Julian Date = Julian Date - 2400000.5} = 56019.28665), $q \equiv {\rm M}_2/{\rm M}_1$ is the system mass ratio, and $V_{COM}$ is the center of mass (systemic) radial velocity in {\kms}.  The primary radial velocity as a function of time $t$, $RV_1(t)$, is
\begin{equation}
RV_1(t) = K_1\left[e\cos{\omega} + \cos{(T(t)+\omega)}\right]+ V_{COM}
\end{equation}
where
\begin{equation}
K_1 = \frac{2\pi{a}\sin{i}}{P\sqrt{1-e^2}} \frac{q}{1+q}
\end{equation}
and the true anomaly $T(t)$ is related to the eccentric anomaly $E(t)$ by
\begin{equation}
\tan{\frac{T(t)}{2}} = \sqrt{\frac{1+e}{1-e}}\tan{\frac{E(t)}{2}}
\end{equation}
which is iteratively solved using Kepler's Equation:
\begin{equation}
M(t) - M_0 = 2\pi\frac{t-{\tau}_0}{P} = E(t) - e\sin{E(t)}.
\end{equation}
These parameters can be used to compute the total system mass (M$_{tot} = a^3/P^2$ in solar masses) and component masses (M$_1$ = M$_{tot}$/[1$+q$], M$_2$ = $q$M$_{1}$). 

We selected an initial parameter set that visually coincided with the primary radial velocity curve through manual experimentation, and enforced the conditions 0.2~yr $\leq P \leq$ 30~yr, $e$ $\leq$ 0.95, 0.005 $\leq q \leq$ 1,  and (initially) M$_{tot}$ $\leq$ 0.3~{\msun}, where the last condition assumes neither primary nor secondary can be more massive than 0.15~{\msun}. 
We then computed a trial MCMC chain of 7$\times$10$^5$ steps, again using the Metropolis-Hastings algorithm, with new parameters drawn from normal distributions with fixed widths $\vec{\beta}$ = (0.5~yr, 0.5~AU, 0.3, 5$\degr$, 5$\degr$, 5$\degr$, 0.2, 2.0~{\kms}). Observed radial velocities were compared to model values calculated at the same epoch using a $\chi^2$ statistic:
\begin{equation}
\chi^2 = \sum_{j=1}^{N_{RV_1}}\frac{(RV_{1}^{(obs)}(t_j)-RV_{1}^{(model)}(t_j))^2}{\sigma_{RV_{1}}^2(t_j)}
\end{equation}
where $N_{RV_1}$ = 13  is the number of primary {\rv} measurements, and $\sigma_{RV_1}$ the measurement errors, each with an additional 0.5~{\kms} systematic error added in quadrature. For a 7-parameter model, this fit had 6 degrees of freedom. 

Following this initial chain, we performed $N_C$ = 20 independent MCMC chains, each encompassing 10$^6$ steps, where the initial parameter set of each chain was chosen from uniform distributions centered on the best-fit model of the initial chain and with half-widths equal to the greater of the trial widths listed above or the standard deviations of the last 75\% of the trial chain (the latter were used for $i$, $\omega$, and $M_0$). These chains were propagated, convergence was verified for all parameters using the \citet{doi:10.1214/ss/1177011136} scale reduction factor, and the last 75\% of all chains were retained for our final distribution. 

As described in the main text, two separate MCMC analyses were performed; one using a weak constraint on the total system mass (M$_{tot}$ $\leq$ 0.30~{\msun}), and a second using constraints based on the spectral composition of the system and the evolutionary models of \citet{2003A&A...402..701B}. The evolutionary models impose two related constraints on the orbit: first, a limit on the total system mass of 0.01~{\msun} $\leq$ M$_{tot}$ $\leq$ 0.14~{\msun}, based on the range of system masses over 0.2--10~Gyr (Figure~\ref{fig:model_masses}); and second, a limit on the mass function of the system:
\begin{equation}
f_M^{(orb)}\sin{i} = K_1\left(\frac{P}{2\pi{G}}\right)^{1/3} \sqrt{1-e^2} \leq  a\frac{q}{1+q}\left(\frac{4\pi^2}{GP^2}\right)^{1/3}. \label{eqn:fmorb}
\end{equation}
Assuming that $f_M^{(orb)}$ cannot exceed $f_M^{(evol)}$ (Equation~\ref{eqn:fmevol}) for the oldest age modeled, we retain only solutions with $f_M^{(orb)} \leq f_M^{(evol)}$(10~Gyr) = 0.25~M$_{\odot}^{1/3}$. This effectively eliminates the $P$ = 3--5~yr minor solutions in the unconstrained fits. The mass function also provides soft 
constraints on the orbital inclination and age of the system.  
Applying the same constraint above as $f_M^{(evol)}$(10~Gyr)$\sin{i} \geq f_M^{(orb)}\sin{i}$ to the left side of Equation~\ref{eqn:fmorb} imposes a minimum value for $\sin{i}$ for a given set of orbital parameters.
Conversely, requiring that $f_M^{(orb)}$ be at least as large as $f_M^{(evol)}$, even for $\sin{i}$ = 1, imposes a minimum constraint on the component masses and hence minimum model age of the system. The distributions of these minimum parameters for all the orbital fits are shown in Figure~\ref{fig:minageinc}. 

To examine predictions for projected separation and primary astrometric perturbation, we combined our 7-parameter model set with the trignometric distance of {\namesh}, $d$ = 23.2$\pm$0.5~pc \citep{2012ApJS..201...19D} to calculate the projected
angular separation vector from primary component to secondary component,  $\vec{\rho}$ = ($\Delta\alpha$(t), $\Delta\delta$(t)). This was determined from
\begin{align}
\Delta\alpha(t) = \frac{a}{d} \left[A(\cos{E(t)} - e) + F\sqrt{1-e^2}\sin{E(t)}\right] \\
\Delta\delta(t) = \frac{a}{d} \left[B(\cos{E(t)} - e) + G\sqrt{1-e^2}\sin{E(t)}\right]
\end{align}
where $\Delta\alpha$ and $\Delta\delta$ are the angular separations on the sky measured in arcseconds, and $A$, $B$, $F$ and $G$ are the Thiele-Innes constants \citep{1907Obs....30..310I,1927BAN.....3..261V}:
\begin{align}
A & = \cos{\omega}\cos{\Omega} - \sin{\omega}\sin{\Omega}\cos{i} \\
B & = \cos{\omega}\sin{\Omega} + \sin{\omega}\cos{\Omega}\cos{i} \\
F & = -\sin{\omega}\cos{\Omega} - \cos{\omega}\sin{\Omega}\cos{i} \\
G & = - \sin{\omega}\sin{\Omega} + \cos{\omega}\cos{\Omega}\cos{i}.
\end{align}
Here, $\Omega$ is the longitude of ascending node, for which we had no constraints, so  a uniform distribution of 0 $\leq \Omega \leq$ 360$\degr$ was assumed. 
The amplitude of astrometric variability was computed from projected angular separation 
\begin{equation}
\rho_{ast} = \rho\left(\mathit{f}_F - \mathit{f}_M\right)
\end{equation}
where 
\begin{equation}
\mathit{f}_F = \frac{f_1}{f_1+f_2} = (1+10^{0.4\Delta{m}})^{-1}
\end{equation}
\begin{equation}
\mathit{f}_M = \frac{M_2}{M_1+M_2}
\end{equation}
is the fractional primary flux, with $\Delta{m} = m_2 - m_1$; and the fractional secondary mass, respectively. To compare to the MKO $J$-band measurements of \citet{2012ApJS..201...19D}, we assumed $\Delta{m}$ = 1.8 based on the spectral template fitting in Section~\ref{sec:spex}.  


\end{document}